\documentclass[%
 reprint,
 amsmath,amssymb,
 aps,
]{revtex4-1}

\usepackage{graphicx}
\usepackage{dcolumn}
\usepackage{bm}

\usepackage{hyperref}
\hypersetup{colorlinks,allcolors=blue}
\usepackage{ulem}

\begin{document}


\title{Atomic Cluster Expansion for a General-Purpose Interatomic Potential of Magnesium}

\author{Eslam Ibrahim}
\email{eslam.saadibrahim@rub.de}
\author{Yury Lysogorskiy}

\author{Matous Mrovec}

\author{Ralf Drautz}
\email{ralf.drautz@rub.de}
\affiliation{ICAMS, Ruhr Universit\"at Bochum, 44780 Bochum, Germany}


\begin{abstract}
We present a general-purpose parameterization of the atomic cluster expansion (ACE) for magnesium. The ACE shows outstanding transferability over a broad range of atomic environments and captures physical properties of bulk as well as defective Mg phases in excellent agreement with reference first-principles calculations. We demonstrate the computational efficiency and the predictive power of ACE by calculating properties of extended defects and by evaluating the P-T phase diagram covering temperatures up to 3000 K and pressures up to 80 GPa. We compare the ACE predictions with those of other interatomic potentials, including the embedded-atom method, an angular-dependent potential, and a recently developed neural network potential. The comparison reveals that ACE is the only model that is able to predict correctly  the phase diagram in close agreement with experimental observations.
\end{abstract}

\maketitle

\section{Introduction}

Magnesium is a lightweight and abundant metal, which makes it an attractive candidate material for automotive and aerospace components~\cite{mordike2001magnesium}. However, its low melting temperature and brittleness, attributed to the hexagonal close-packed (hcp) structure, limit the range of possible applications~\cite{wu2015origins}. The hcp crystal structure of Mg as well as the body-centered cubic (bcc) structure of Na and the face-centered cubic (fcc) structure of Al, its neighbors in the periodic table, can be understood from the nearly free electron approximation~\cite{pettifor1995bonding}. In all these simple metals, the atomic cores induce Friedel oscillations in the electron density with characteristic wave lengths that depend on the Fermi energy and govern the stabilization of the bcc, hcp and fcc crystal structures. Changes of the electronic structure under compression lead to various phase transitions~\cite{pettifor1995bonding, Grimvall_RevModPhys.84.945}. While the phase diagram of Mg has been investigated both theoretically~\cite{mcmahan1983structural, mcmahan1986pressure, moriarty1995first, mehta2006ab, li2010crystal, Cui2022} and experimentally~\cite{olijnyk1985high, errandonea2001melting, Errandonea_2003, Stinton_PhysRevB.90.134105, Beason_PhysRevB.104.144106}, exact locations of phase boundaries and existence of some phases are still under debate.

Atomistic simulations are nowadays indispensable for a detailed exploration of mechanical and thermodynamic properties. There exist several interatomic potentials for Mg, including embedded-atom method (EAM) potentials \cite{liu1998grain, zhou2004misfit, sun2006crystal, wilson2016unified, pei2018tunable},  modified embedded-atom method (MEAM) potentials \cite{kim2009atomistic, dickel2018new, ahmad2020analysis}, angular-dependent potentials (ADP) \cite{smirnova2018new}, tight-binding models \cite{cleri1993tight, li2015gupta}, and neural-network potentials (NNPs) \cite{stricker2020machine, dickel2021lammps} and a moment tensor potential (MTP) \cite{Poul23}.  However, a recent study~\cite{troncoso2022evaluating} showed that none of the potentials that were considered was able to predict accurately the $P$-$T$ phase diagram even for intermediate ranges of temperatures and pressures. 

Here we present an interatomic potential for Mg that is able to describe reliably very different atomic configurations and is applicable to large-scale atomistic studies of mechanical as well as thermodynamic properties. The excellent transferability is achieved by parameterizing the atomic cluster expansion (ACE)~\cite{drautz2019atomic} based on a wide range of density functional theory (DFT) reference data. We validate the potential in detail and demonstrate an excellent match to the reference data, in particular for properties that were not part of the training. We then apply the model to predict the Mg phase diagram.

The paper is structured as follows. In Sec.~\ref{sec:methods} we briefly review theoretical methods that we employ for the construction of ACE for Mg and for computations of mechanical and thermodynamic properties. In Sec.~\ref{sec:ace} we describe the ACE parameterization procedure. Sec.~\ref{sec:validation} contains validation studies for fundamental structural, elastic and vibrational properties. In Sec.~\ref{sec:sf} we present the calculations of stacking faults and in Sec.~\ref{sec:phasediag} we provide a comprehensive study of thermodynamic properties of Mg and evaluate the $P$-$T$ phase diagram using state-of-the-art thermodynamic integration techniques. Conclusions are provided in Sec.~\ref{sec:conclusion}.

\section{Methods} \label{sec:methods}

\subsection{DFT reference calculations}
We employ the all-electron FHI-aims code~\cite{fhi-aims1,fhi-aims2} to carry out first-principles reference calculations based on density functional theory (DFT). All DFT calculations were performed using the PBE functional \cite{perdew1996generalized}, tight basis settings, a k-mesh density of 0.175 per \AA$^{-1}$, and Gaussian smearing of 0.1\,eV.

\subsection{Atomic cluster expansion}
The ACE parameterization was fitted using the package Pacemaker \cite{bochkarev2022efficient}. We employed a Finnis-Sinclair-type mildly non-linear representation of the atomic energy that incorporates two atomic properties that are represented by linear, in principle complete, ACE basis expansions
\cite{drautz2019atomic,lysogorskiy2021performant,bochkarev2022efficient}. The non-linear representation can be motivated from the second-moment approximation \cite{qamar2022atomic} and was shown to be efficient \cite{bochkarev2022efficient} for metals \cite{lysogorskiy2021performant} as well as covalently bonded materials \cite{qamar2022atomic,bochkarev2022multilayer}. A detailed description of the ACE methodology can be found in original  Refs.~\cite{drautz2019atomic, Dusson22, drautz2020atomic, lysogorskiy2021performant, bochkarev2022efficient}.

\subsection{Molecular dynamics and statics simulations}
The Large-scale Atomic/Molecular Massively Parallel Simulator (LAMMPS) \cite{LAMMPS} and the Performant implementation of the atomic cluster expansion (PACE) \cite{lysogorskiy2021performant} were used to carry out molecular statics and molecular dynamics (MD) simulations. Some calculations for other potentials were performed using the Atomic Simulation Environment (ASE) \cite{larsen2017atomic}.

\subsection{Computation of free energy}
Non-equilibrium thermodynamic integration was used for free energy calculations as implemented in the software package CALPHY \cite{menon2021automated}. Two different integration paths were employed: the Frenkel-Ladd path \cite{frenkel1984new} for computing the free energy at a given temperature and volume, and reversible scaling \cite{de1999optimized, de2001single} for generating the free energy for a range of temperatures. We carefully converged the simulations to ensure free energy difference errors to be significantly smaller than 1\,~meV (see \cite{suppl}).


\section{Parameterization}
\label{sec:ace}

\begin{figure*}[hbt!]
\centering
\includegraphics[width=15cm,height=10cm]{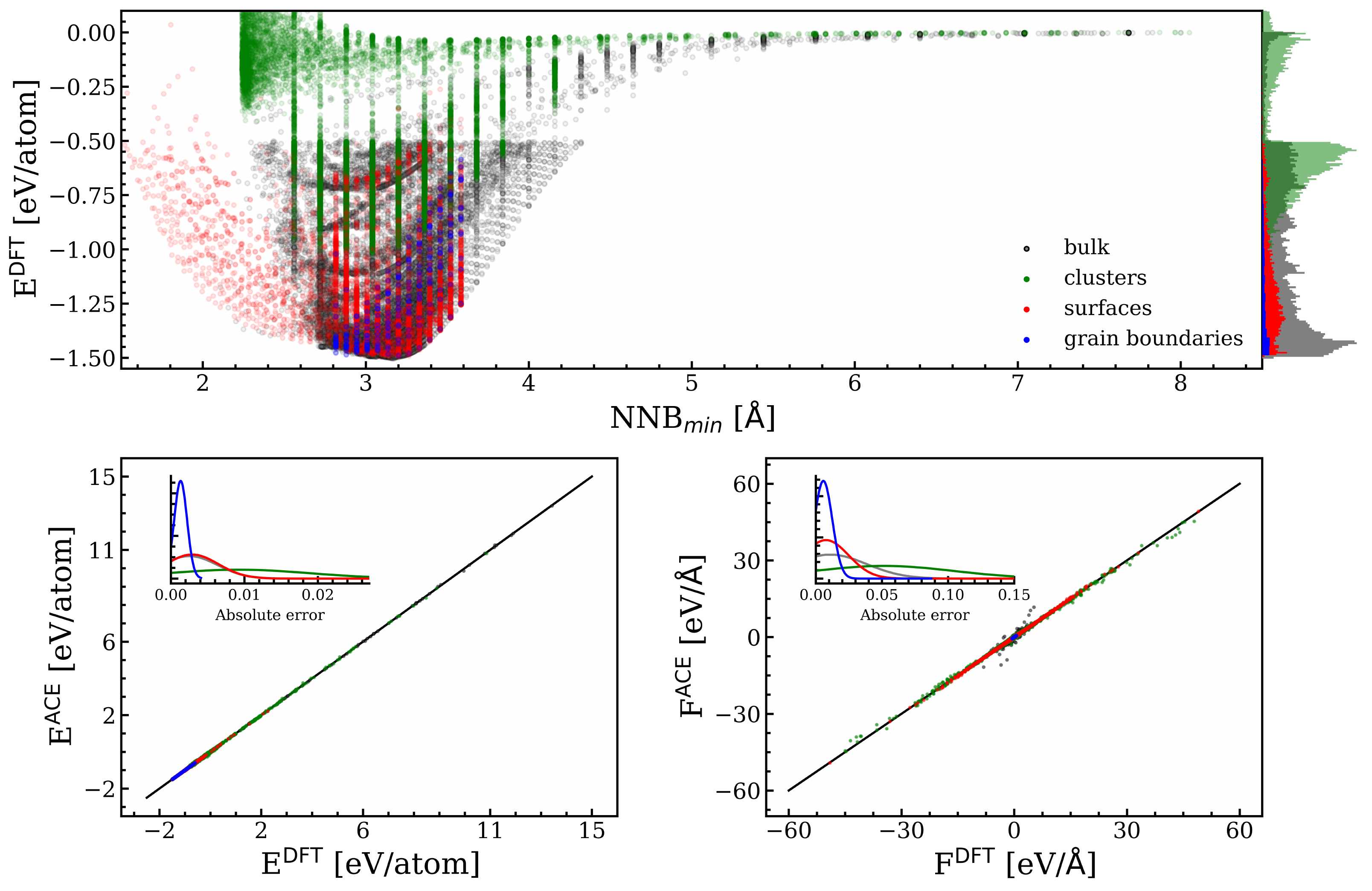}
\caption{DFT energies for the training dataset plotted with respect to the nearest neighbor distance in each structure (top panel). Cross-correlation plots and error distributions for energy and force showing an excellent agreement between ACE and DFT (bottom graphs).}
\label{train_data}
\end{figure*}

We carried out more than 120 thousand individual DFT reference calculations to sample local atomic environments as exhaustively as possible. The workflow management software pyiron \cite{janssen2019pyiron} was employed to automatize some calculations. The atomic configurations comprised of different bulk structures including supercells with displaced atoms, point defects, planar defects and surfaces. In addition, we included small random clusters containing 2\, to 5\, atoms to ensure transferability to a large variety of atomic environments. We did not include any explicit information on liquid phases. For the bulk phases, we sampled a range of volumes such that the nearest neighbor distances  varied between about 0.7\, and 3.0\, of the nearest neighbor distance of the equilibrium hcp phase. The reference DFT data are plotted with respect to the nearest neighbor distance in each structure in the top panel of Fig.~\ref{train_data}.

For the parameterization, we employed a successive hierarchical basis extension with power-order ranking \cite{bochkarev2022efficient} and a cutoff of  8.2\,\AA. From the full DFT reference dataset, we selected 40000 structures for training (corresponding to 738705\, atoms) and 5000 structures for testing (corresponding to 92900\, atoms). The final ACE parametrization contained 724\, basis functions and a total of 2528 parameters. Energies and forces in the structures were weighted using an energy based weighting scheme \cite{bochkarev2022efficient} that gave higher weights to low energy structures. Specifically, 75\% of the weight was given to structures that were within 1\,eV from the lowest energy structure.


Cross-correlation plots in Fig.~\ref{train_data} show that ACE reproduces the DFT reference data with excellent accuracy. The training metrics for the complete training set gives a mean-absolute error of 5.4\, meV/atom for the energy and 31.5\, meV/\AA\ for the force. Corresponding errors for the test set of 5.7\, meV/atom and 31.9 meV/\AA, respectively, demonstrate that the model is not overfitted. Structures that are within 1\, eV from the ground state were given the largest weight and there errors are 4.7\, meV/atom and 29.4\, meV/\AA\ for train and 4.9\, meV/atom and 29.8\, meV/\AA\ for test. Additional details of the training and error analysis are given in the supplemental material~\cite{suppl}.

\section{Validation}
\label{sec:validation}

We validated the ACE parametrization by computing a number of materials properties. In addition, we compared ACE predictions with those of three other potentials for Mg, namely, an EAM potential developed for studies of solid-liquid interfaces~\cite{wilson2016unified}, an ADP parametrization for pure Mg and the Mg-H system~\cite{smirnova2018new}, and a recent NNP model aimed at studies of extended defects~\cite{stricker2020machine}.

\subsection{Equilibrium structure and elastic moduli}

The equilibrium lattice constant $a$ and the $c/a$ ratio of the hcp phase at 0 K are reproduced by all potentials in close agreement with DFT and experiment (see Table~\ref{elmat_table}).  The same holds for the elastic constants that are also summarized in Table~\ref{elmat_table}.

\begin{table}[hbt!]
\begin{center}
\caption{Basic materials properties of the Mg hcp phase predicted by different interatomic potentials, DFT and experiment. The transformation pressure $P_{tr}$ between the hcp and bcc phases and the melting temperature $T_{m}$ of the hcp phase at 0 K are also included.}
\label{elmat_table}
  \begin{tabular}{lccccc|c}
  \toprule
    Property   & DFT$^{\text{\cite{sin2009ab}}}$& ACE &EAM$^{\text{\cite{wilson2016unified}}}$ & ADP$^{\text{\cite{smirnova2018new}}}$& NNP$^{\text{\cite{stricker2020machine}}}$ & Expt$^{\text{\cite{slutsky1957elastic}}}$ \\
 \hline
  $a$ [\AA]        & 3.190 & 3.187 & 3.185 & 3.188 & 3.186  & 3.209  \\
  $c/a$            & 1.627 & 1.623 & 1.628 & 1.633 & 1.613 & 1.624  \\
  $C_{11}$ [GPa] & 63.44 & 71.48 & 70.42 & 53.32 & 72.42 & 63.48  \\
  $C_{12}$ [GPa] & 26.15 & 24.30 & 25.56 & 16.74 & 30.31 & 25.94  \\
  $C_{13}$ [GPa] & 21.07 & 20.93 & 15.44 & 15.67 & 26.73 & 21.70  \\
  $C_{33}$ [GPa] & 68.47 & 67.14 & 69.05 & 73.62 & 68.22 & 66.45  \\
  $C_{44}$ [GPa] & 18.32 & 18.12 & 12.68 & 14.93 & 19.10 & 18.42  \\
  $P_{tr}$ [GPa] & 52.5\footnote{We calculated this value using FHI-aims.}  & 51.6 & 34.4 & 23.7 & 5.5 & 
  $\approx$ 50$^{\text{\cite{olijnyk1985high}}}$ \\
  $T_{m}$ [K]    & 910 & $\approx$ 900  & 918  & 910 & $\approx$ 900 &  923 \\
 \toprule
  \end{tabular}
\end{center}
\end{table}

\subsection{Structural stability and transition pressure}

\begin{figure*}[hbt!]
\centering
\includegraphics[width=\textwidth]{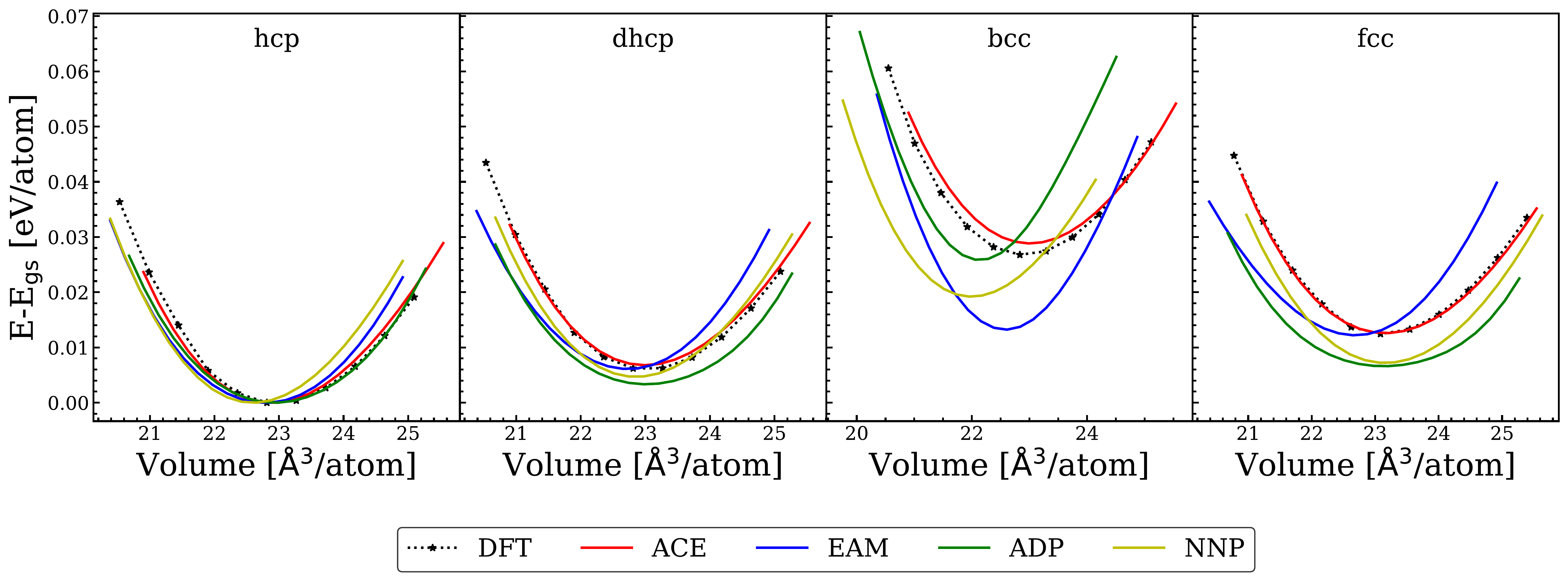}
\caption{Volume dependence of the energy for the most relevant Mg phases computed using ACE, EAM, ADP, NNP and DFT. For each method, the energy of its equilibrium hcp structure was subtracted.}
\label{fig:phases}
\end{figure*}

In Fig.~\ref{fig:phases}, we compare the structural energies of the most relevant Mg phases -- hcp, dhcp, bcc and fcc -- computed using ACE, EAM, ADP, NNP and DFT. ACE captures the structural energy differences in quantitative agreement with DFT for all considered phases while the other potentials show significant discrepancies for the bcc phase. 

\begin{figure*}[hbt!]
\centering
\includegraphics[width=\textwidth]{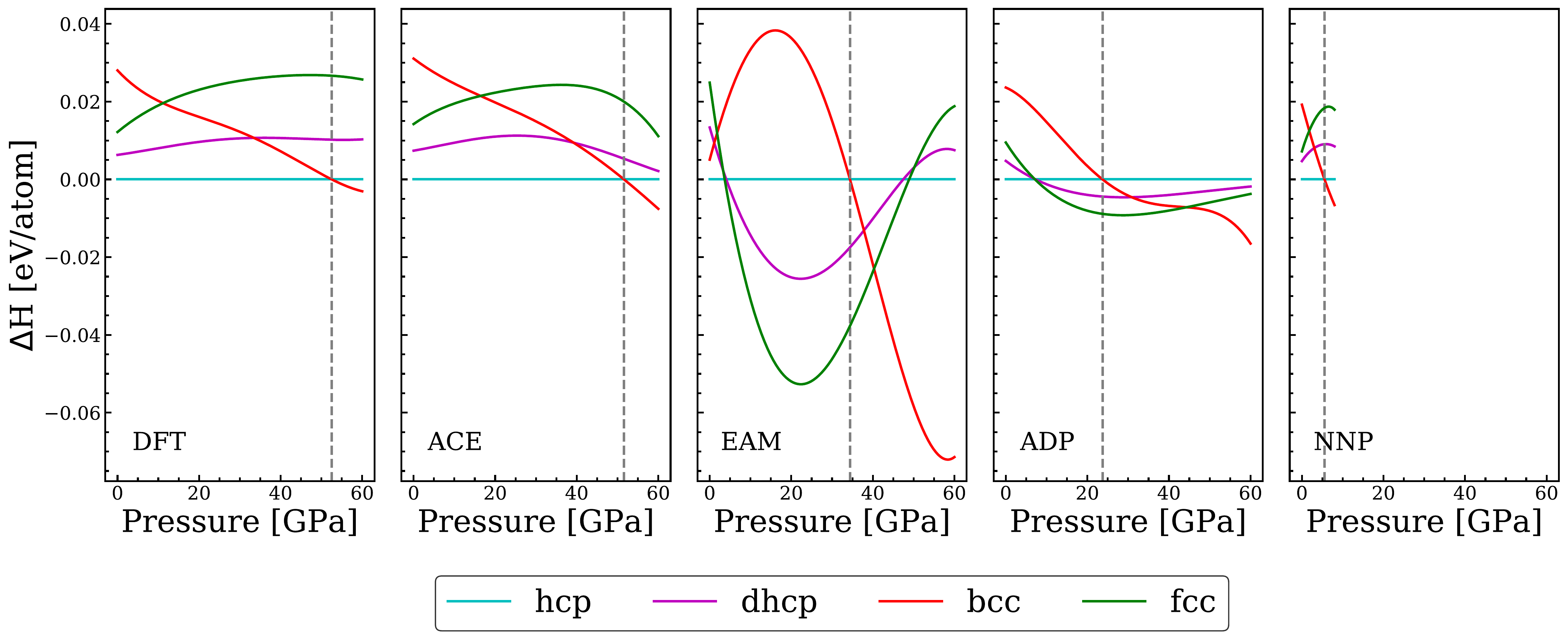}
\caption{Relative enthalpies (at 0\, K) of the dhcp, bcc and fcc phases with respect to that of the hcp phase as a function of pressure computed using ACE, EAM, ADP, NNP and DFT.}
\label{fig:dft_dh}
\end{figure*}

The energy difference between hcp and bcc is critical for the hcp-bcc phase transition. Figure~\ref{fig:dft_dh} shows relative enthalpies of the bcc, fcc and dhcp phases with respect to that of the hcp phase as a function of pressure.  It is apparent that only ACE agrees well with the DFT predictions while the other three potentials show not only quantitatively but even qualitatively incorrect behavior. The theoretically predicted transition pressures $P_{tr}$ from hcp to bcc at zero temperature (marked by vertical dashed lines in Fig.~\ref{fig:dft_dh}) are 52.5\,, 51.6\,,  34.4\,, 23.7\, and 5.5\, GPa for DFT, ACE, EAM, ADP and NNP, respectively. Our DFT and ACE values fall into the range of 47-55 GPa obtained by other electronic structure~\cite{moriarty1995first, mehta2006ab, sin2009ab} and experimental \cite{olijnyk1985high, Stinton_PhysRevB.90.134105, Beason_PhysRevB.104.144106} studies. Apart from quantitative discrepancies,  EAM and ADP predict Mg to transform first to the fcc phase. The NNP model was trained mostly on structures close to the equilibrium densities and therefore is unable to predict correctly the behavior at elevated pressures.

\subsection{Phonon spectra}

\begin{figure*}[hbt!]
\centering
\includegraphics[width=15cm, height=10cm]{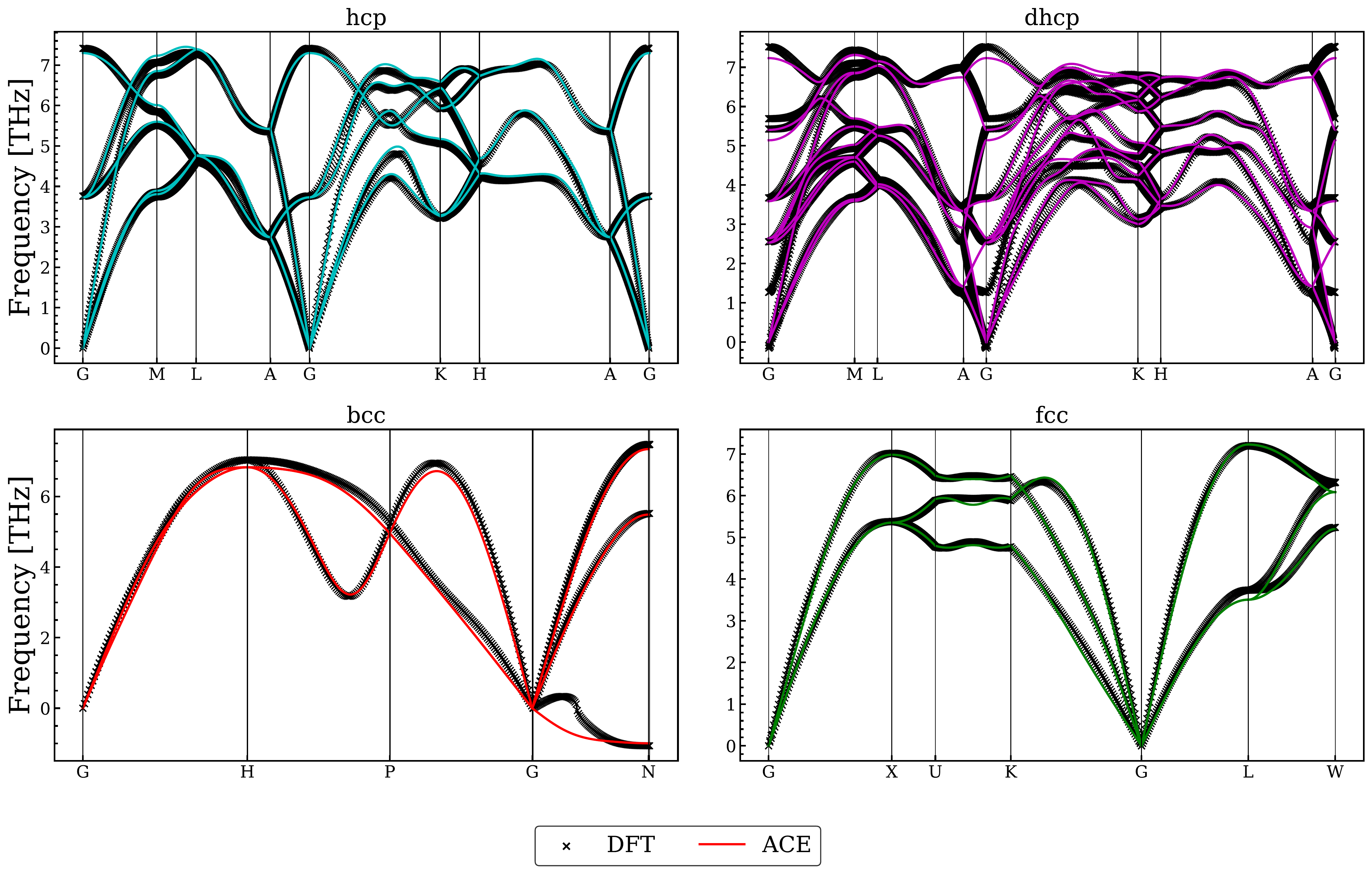}
\caption{Phonon band structures for hcp, dhcp, bcc, and fcc computed with ACE and DFT.} 
\label{phonon_bands}
\end{figure*}

In Fig.~\ref{phonon_bands}, we present the computed phonon band structures for hcp, dhcp, bcc and fcc using ACE and DFT. We calculated the phonon spectra using the force constants method as implemented in the Phonopy package \cite{togo2015first}. We observe very close agreement between ACE and DFT, including the prediction of phonon softening in bcc. The phonon band structures for EAM, ADP and NPP show significant deviations from DFT~\cite{suppl}.

\subsection{Surface energies}

We calculated energies of various surfaces of Mg in the hcp and bcc phases, as displayed in Table~\ref{surfs_ene}. The atomic configurations and DFT data were taken from the Crystalium database \cite{kirklin2015open, tran2019anisotropic, zheng2020grain}. The surfaces of the hcp phase were not part of the training reference data. Hence, the prediction of surface energies presents a stringent test of the ACE transferability.  For hcp, the close-packed (0001) surface has the lowest energy with a deviation between ACE and DFT of only 2\, meV/\AA$^{2}$. Other surfaces are reproduced equally well or with only slightly larger errors. 

 \begin{table}[hbt!]
\begin{center}
\caption{Calculated surface energies (in eV/\AA$^{2}$) for hcp and bcc phases. The DFT values are taken from the Crystalium database \cite{kirklin2015open, tran2019anisotropic, zheng2020grain}.}
\label{surfs_ene}
  \begin{tabular}{lccclcc}
  \toprule
    hcp  &  ACE  & DFT  & & bcc &  ACE & DFT\\
 \hline
  (0001)                & 0.034 & 0.032  & & (111)     & 0.041 & 0.047 \\
  ($10\bar{1}0$)        & 0.036 & 0.037  & & (211)     & 0.028 & 0.030 \\
  ($10\bar{1}1$)        & 0.047 & 0.039  & & (321)     & 0.028 & 0.030 \\
  ($21\bar{3}0$)        & 0.040 & 0.044  & & (110)     & 0.033 & 0.037 \\
  ($10\bar{1}2$)        & 0.049 & 0.044  & & (320)     & 0.048 & 0.038 \\
  ($11\bar{2}0$)        & 0.041 & 0.045  & & (210)     & 0.026 & 0.039 \\
  ($21\bar{3}1$)        & 0.040 & 0.046  & & (310)     & 0.033 & 0.039 \\
  ($21\bar{3}2$)        & 0.050 & 0.046  & & (331)     & 0.041 & 0.041 \\
  ($2\bar{1}\bar{1}2$)  & 0.053 & 0.046  & & (322)     & 0.050 & 0.043 \\
  ($22\bar{4}1$)        & 0.041 & 0.047  & & (311)     & 0.042 & 0.044 \\
  ($11\bar{2}1$)        & 0.039 & 0.047  & & (100)     & 0.058 & 0.044 \\
  ($20\bar{2}1$)        & 0.035 & 0.048  & & (221)     & 0.063 & 0.047 \\
   &  &                                 & & (332)     & 0.055 & 0.048 \\
 \toprule
  \end{tabular}
\end{center}
\end{table}

\subsection{Transformation paths}

\begin{figure*}[hbt!]
\centering
\includegraphics[width=15cm, height=10cm]{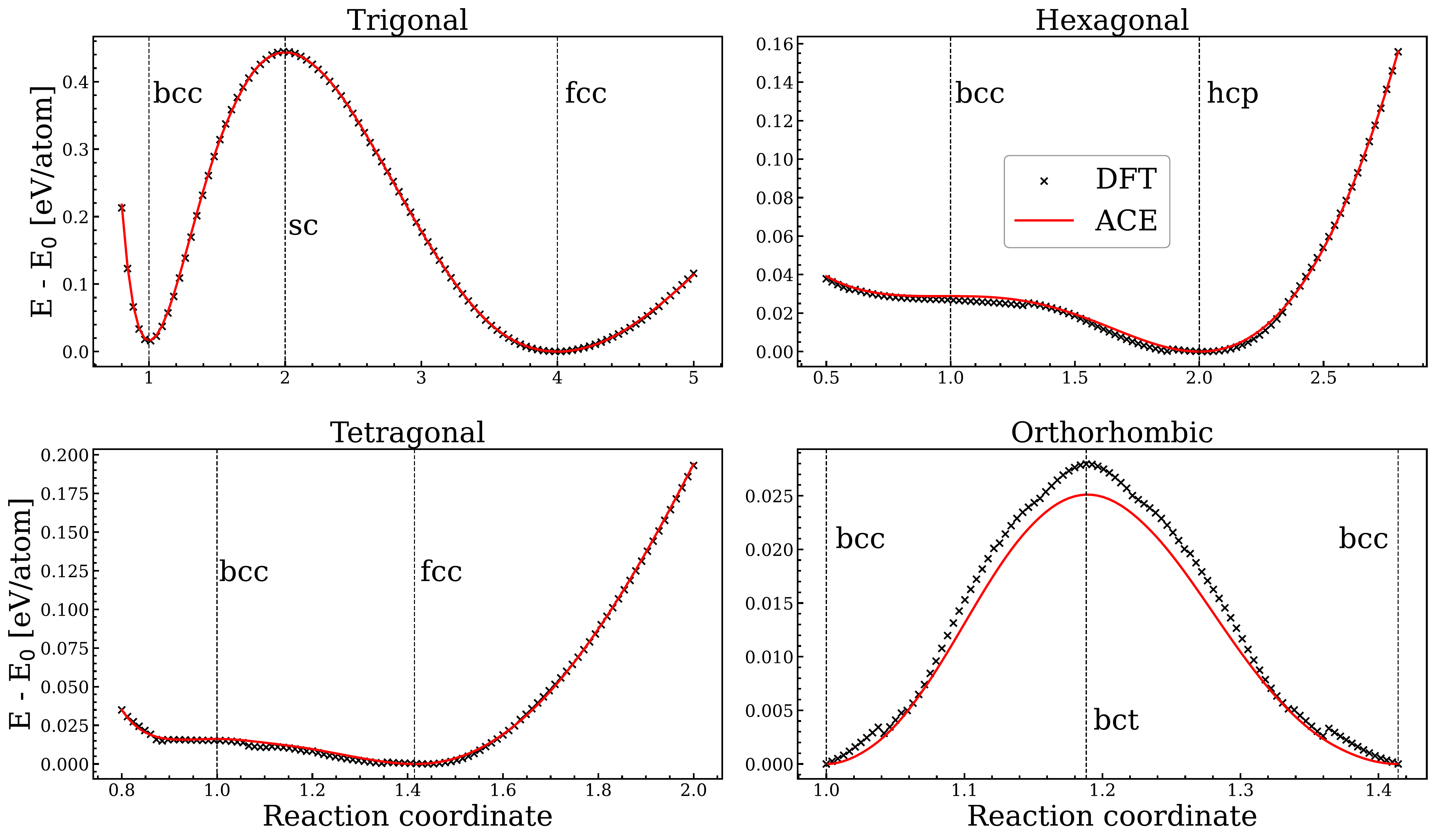}
\caption{Trigonal, hexagonal, tetragonal, and orthorhombic transformation paths for ACE and DFT.} 
\label{tp}
\end{figure*}

Along transformation paths between crystal structures coordination and bond distances change significantly. To illustrate the capability of ACE to describe large bonding rearrangements, we calculated the energy along four different transformation paths. In Fig.~\ref{tp}, we show the trigonal, hexagonal, tetragonal, and orthorhombic paths as predicted by ACE and DFT. The discontinuities that are visible in the DFT data of the orthorhombic path are due to $k$-point re-meshing. Results for the other potentials are provided in the Supplemental material~\cite{suppl}.

\section{Stacking faults and generalized stacking fault energy surfaces}
\label{sec:sf}

The low ductility of Mg at ambient temperatures is related to limited number of available slip systems \cite{mordike2001magnesium, rohrer2001structure}. The close-packed basal plane offers two independent primary slip systems where the $\langle a \rangle$ dislocations glide at relatively low stresses \cite{hutchinson2010effective, sanchez2014measuring}. To initiate plasticity during loading along the $\langle c \rangle$ axis necessitates glide of $\langle c+a \rangle$ dislocations on secondary pyramidal slip systems. This usually requires a thermal activation and elevated stresses \cite{mordike2001magnesium, fan2015towards, geng2014structure, wu2015origins, Itakura2016, xie2016pyramidal}. 

\begin{figure*}[hbt!]
\centering
\includegraphics[width=15cm,height=10cm]{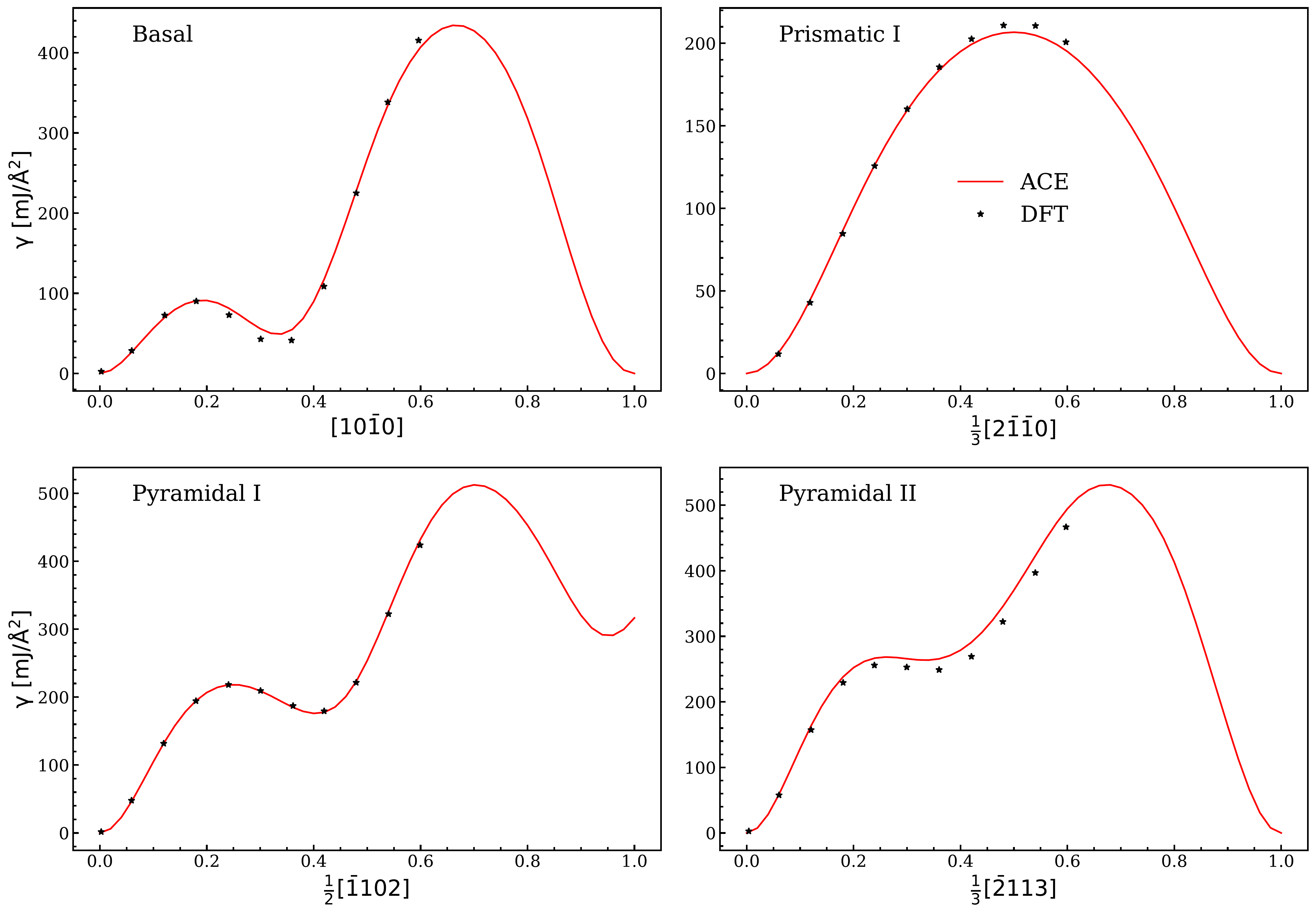}
\caption{Generalised stacking fault energy $\gamma$-lines on several planes in hcp Mg. Panels show data for the basal,  prismatic I, pyramidal I, and pyramidal II planes from ACE and DFT ~\cite{yin2017comprehensive}.}
\label{gl}
\end{figure*}

\begin{figure*}[hbt!]
\centering
\includegraphics[width=15cm,height=10cm]{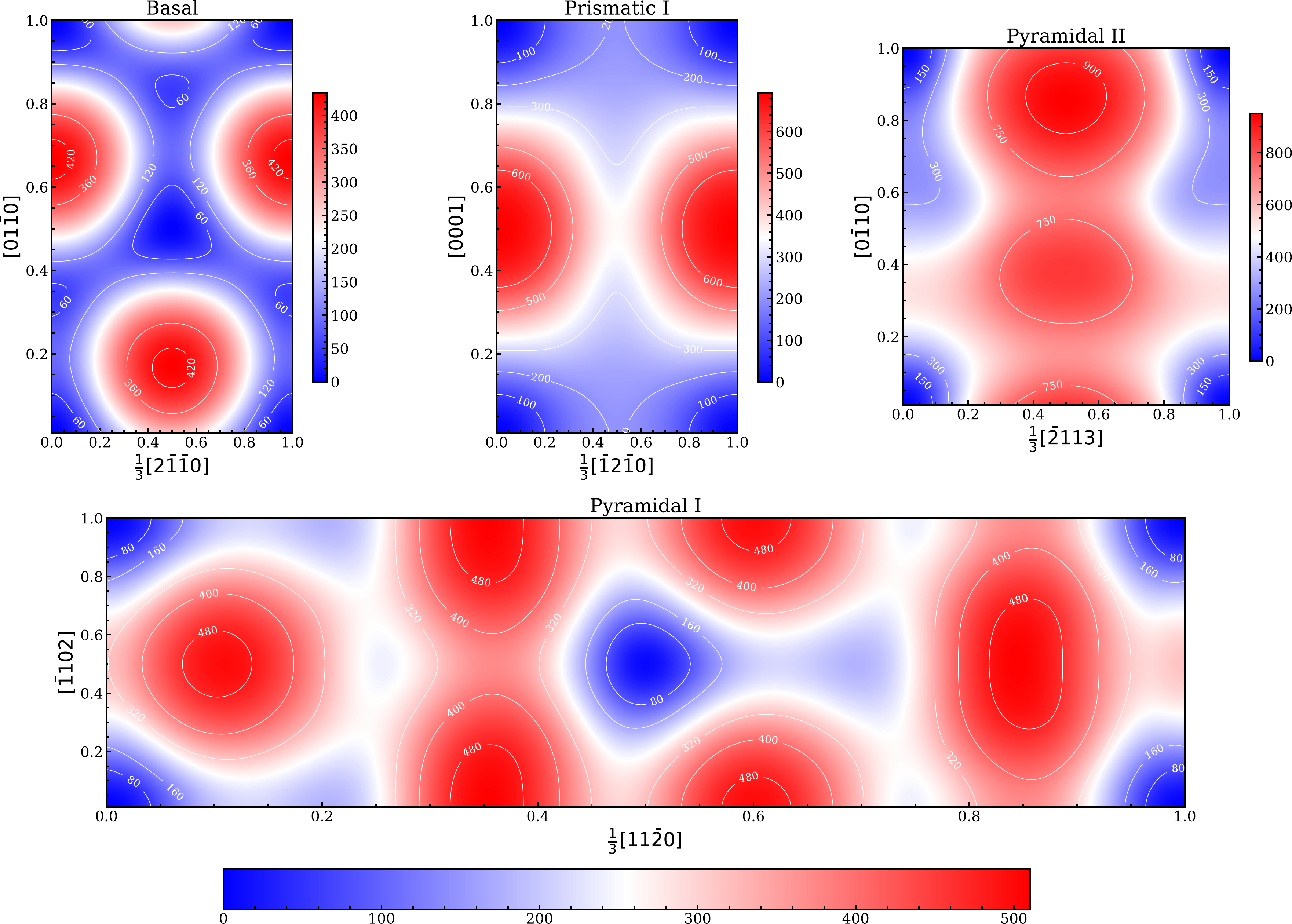}
\caption{The generalised stacking fault energy ($\gamma$-surface) on several planes in hcp Mg. Figures show the data for the basal,  prismatic I, prismatic II,
pyramidal I, and pyramidal II planes calculated using ACE.}
\label{gs}
\end{figure*}

Generalized stacking fault energy (GSFE) surfaces, also known as $\gamma$-surfaces, \cite{vitek1968intrinsic} are an important tool for assessment of dislocation behavior. Since they are accessible by first principles calculations \cite{chetty1997stacking, smith2007surface, shin2011orbital, wen2009systematic, wang2010first, zhang2013first, yin2017comprehensive} and experiments \cite{smallman1970stacking, sastry1969stacking, couret1985situ}, they can be used also for validation of interatomic potentials. It has been reported \cite{yasi2009basal, shin2011orbital} that there exist notable differences in GSFE values obtained by DFT and EAM potentials.

We calculated energies of stable stacking faults  on the basal $\{0001\}$, prismatic I $\{10\bar{1}0\}$, prismatic II $\{11\bar{2}0\}$, pyramidal I $\{10\bar{1}1\}$, and pyramidal II $\{11\bar{2}2\}$ planes and compared them to results of a recent DFT study \cite{yin2017comprehensive}. The comparison in Table~\ref{gamma_table} shows a good agreement for all stacking faults with errors not exceeding 30\, mJ/m$^{2}$. It is important to note that none of the faults was included in the training data.

\begin{table}[hbt!]
\begin{center}
\caption{Stacking fault (SF) energies (in mJ/m$^{2}$) calculated by ACE and DFT~\cite{yin2017comprehensive}. The acronyms ISF and ESF correspond to intrinsic and extrinsic faults, respectively; the letters N and W mark narrow and wide atomic planes (see text); for a detailed description of SF configurations see Ref.~\cite{yin2017comprehensive}.}
\label{gamma_table} 
  \begin{tabular}{lcc}
\toprule
    & ACE & DFT   \\
\hline
Basal & & \\
  ~ISF1           & 25 & 18 \\
  ~ISF2           & 49 & 34 \\
  ~ESF            & 72 & 54 \\
Prismatic I & & \\
  ~W-SF1      & 201 & 212 \\
  ~W-SF2      & 363 & 383 \\
Prismatic II & & \\
  ~SF1       & 206 & 183 \\
Pyramidal I & & \\
  ~N-SF1      & 169 & 165 \\
  ~W-SF1      & 285 &  - \\
  ~W-SF2      & 158 & 161 \\
  ~W-SF3      & 232 & 203 \\
Pyramidal II & & \\
  ~SF1       & 185 & 165 \\
 \toprule
  \end{tabular}
\end{center}
\end{table}

Apart from the stable stacking faults,  we calculated entire $\gamma$-surfaces for the basal $\{0001\}$, prismatic I $\{10\bar{1}0\}$, prismatic II $\{11\bar{2}0\}$, pyramidal I $\{10\bar{1}1\}$, and pyramidal II $\{11\bar{2}2\}$ planes. In Fig.~\ref{gl}, we show first cross sections of the $\gamma$-surfaces along high symmetry directions that are relevant for splitting of dislocation cores. The plots contain also DFT results from Ref.~\cite{yin2017comprehensive}.  Overall, ACE shows an excellent agreement with DFT for all investigated directions. From the cross section on the basal plane one can see that not only the stable but also the unstable stacking fault energy, corresponding to the shift of about $\frac{1}{6} [ 1 0 \bar{1} 1 ]$, is reproduced very accurately and is consistent with existing DFT data~\cite{smith2007surface, yasi2009basal}. The same applies also for both pyramidal planes.

Contour plots of the entire $\gamma$-surfaces are shown in Fig.~\ref{gs}. The $\gamma$-surfaces for the basal and prismatic planes agree well with available DFT results \cite{wen2009systematic}. A single metastable stacking fault (I2) is visible on the basal plane and none on the prismatic planes. 

For the first-order pyramidal I $\{10\bar{1}1\}$ plane, there exist two sets of atomic layers with narrow (N) and wide (W) spacings between them. Here we evaluated the $\gamma$-surface between the widely spaced planes that is relevant for the splitting of the $\langle c+a \rangle$ dislocations. The shapes of of both pyramidal $\gamma$-surfaces as well as the locations of the stacking faults agree quantitatively with values reported in recent DFT studies on Mg~\cite{yin2017comprehensive, Itakura2016, ghazisaeidi2014first} and are also consistent with DFT results for hcp metals Ti~\cite{ready2017stacking} and Zr~\cite{chaari2014first}.

\section{Phase diagram}
\label{sec:phasediag}

\begin{figure*}[hbt!]
\centering
\includegraphics[width=16cm]{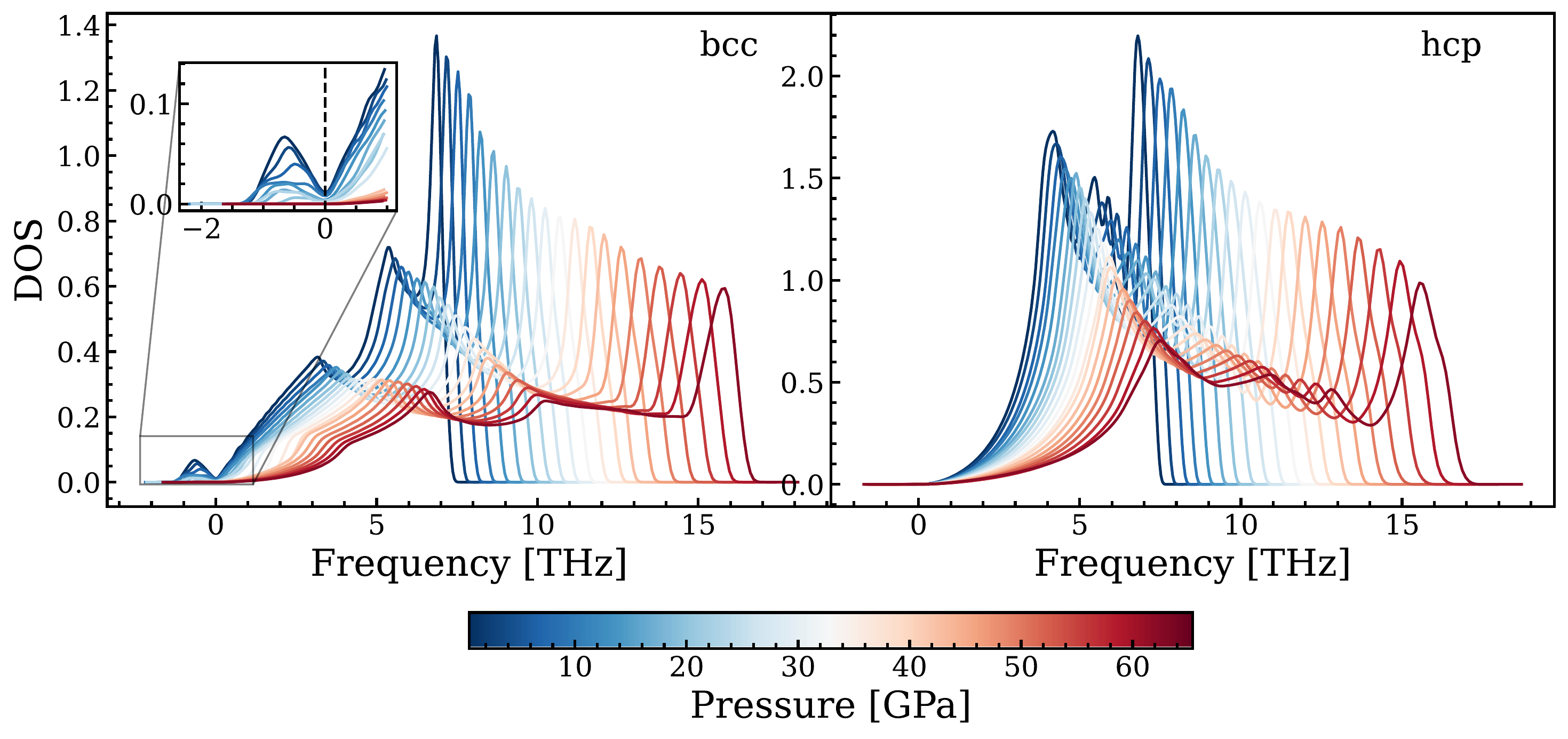}
\caption{Phonon densities of states of the hcp and bcc phases at different pressures as predicted by ACE.}
\label{phon_p}
\end{figure*}

\begin{figure*}[hbt!]
\centering
\includegraphics[width=13cm,height=8cm]{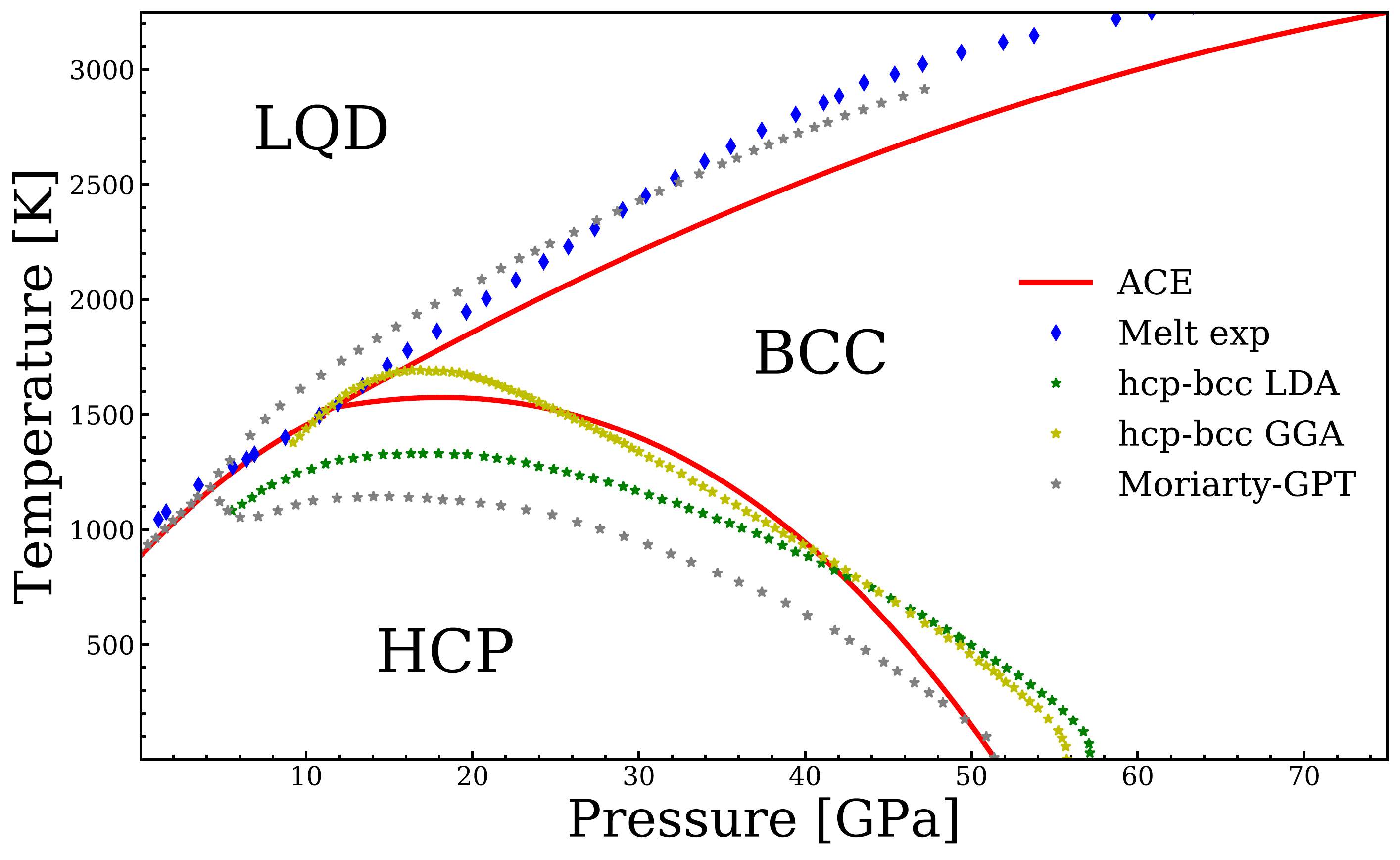}
\caption{Mg phase diagram predicted by ACE together with results from literature from experimental  ~\cite{errandonea2001melting}, as well as theoretical studies  ~\cite{mehta2006ab, moriarty1995first}.}
\label{pd}
\end{figure*}

An evaluation of phase diagram using DFT calculations is computationally expensive so that accurate and efficient interatomic potentials are largely beneficial for such a task. However, a recent comparative study \cite{troncoso2022evaluating} revealed that none of the potentials considered, including several EAM  \cite{liu1998grain, zhou2004misfit, sun2006crystal, wilson2016unified, pei2018tunable} and MEAM \cite{kim2009atomistic, dickel2018new, ahmad2020analysis} potentials, ADP \cite{smirnova2018new},  tight-binding models \cite{cleri1993tight, li2015gupta}, and NNPs \cite{stricker2020machine, dickel2021lammps}, were able to provide a satisfactory description of thermodynamic properties of Mg over a broad range of temperatures and pressures.

We computed the $P$-$T$ phase diagram using non-equilibrium thermodynamic integration (NETI) \cite{freitas2016nonequilibrium}, where we employed two thermodynamic paths: the Frenkel-Ladd path \cite{frenkel1984new} and the reversible-scaling path \cite{de1999optimized, de2001single}. For the automated calculations of the Gibbs free energies of all phases we used the software package CALPHY \cite{menon2021automated}. Examples of the Gibbs free energies at zero pressure computed using NETI for different temperatures together with additional convergence studies are provided in the Supplemental material~\cite{suppl}.

The complete phase diagram predicted by ACE is presented in Fig.~\ref{pd}. For the investigated range of temperatures and pressures, there exist only three phases, namely, hcp, bcc and liquid.  It should be noted that the DFT training data did not contain any configurations at extreme densities and energies so that the present ACE parametrization is reliable up to pressures of about 70 GPa and temperatures up to about 4000 K only.

\begin{figure}[hbt!]
\centering
\includegraphics[width=\columnwidth]
{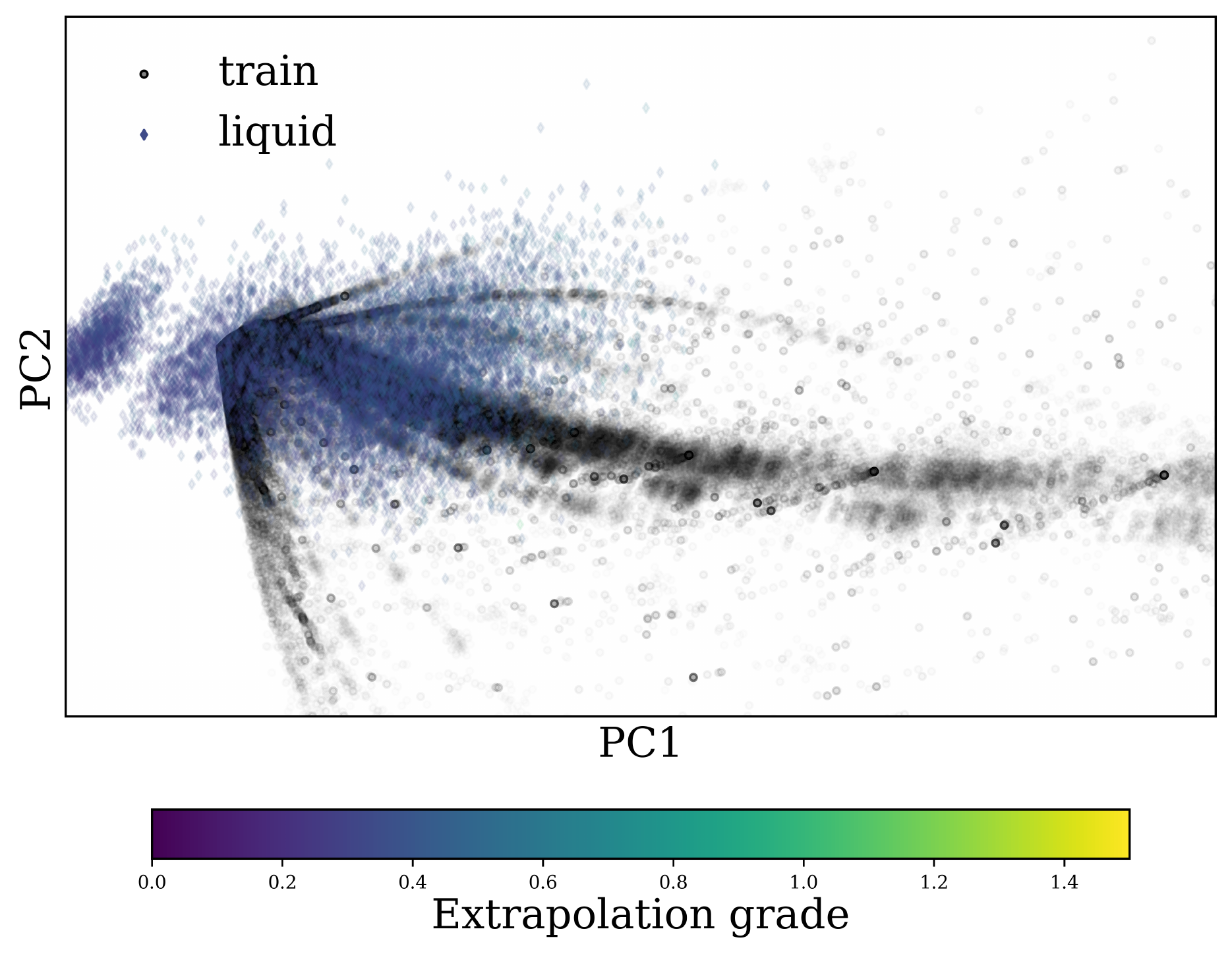}
\caption{PCA of the B-basis function projections for training set and liquid local atomic environments, which were not part of the training.}
\label{pca}
\end{figure}

The liquidus line predicted by ACE shows a good agreement with the experimental data \cite{errandonea2001melting, Cui2022, Stinton_PhysRevB.90.134105, Beason_PhysRevB.104.144106}  as well as recent AIMD results~\cite{Cui2022} despite the fact that no liquid reference DFT data was included in the training procedure. In order to validate that the liquid environments are captured reliably by ACE, we computed the uncertainty indicator $\gamma$ \cite{lysogorskiy2022active} to estimate the extrapolation grade in the description of the local atomic environments occurring in the liquid phase.  A value of $\gamma$ between zero and one indicates interpolation, i.e., reliable prediction, while $\gamma>1$ corresponds to extrapolation and increasingly uncertain predictions. Furthermore, we carried out principal component analysis (PCA) of the ACE basis function for the training configurations as well as for the liquid configurations obtained during the phase diagram simulations up to 13 GPa.  In Fig.~\ref{pca}, we show first two components of PCA decomposition together with the extrapolation grade obtained for the liquid structures. It can be seen that the extrapolation grade is smaller than one for most liquid configurations and that the liquid local atomic environments overlap with the environments in the training dataset. This analysis demonstrates that atomic configurations found in the liquid were efficiently sampled during training, despite the fact that explicitly liquid configurations, for example, from AIMD, where not part of the training set.


The region of the phase diagram in the vicinity of the triple point is still rather uncertain. There exist evidence of additional diffraction peaks~\cite{errandonea2001melting, Stinton_PhysRevB.90.134105} that do not belong to hcp, dhcp or bcc phases, but no other crystalline phase could be identified so far. According to recent shock-release experiments, the triple point of polycrystalline Mg lies at about 20\, GPa and 1650\, K \cite{Beason_PhysRevB.104.144106} which is higher than most theoretical estimates.

To analyze in more detail the origins of the hcp-bcc transformation, we evaluated the phonon density of states for both phases as a function of pressure, as shown in Fig.~\ref{phon_p}.  We find  that hcp is dynamically stable at all pressures while the bcc phase  is unstable below 30 GPa, as evidenced by the negative phonon frequencies.  The dynamical instability of the bcc phase at low pressures is consistent with available DFT results~\cite{Wentzcovitch_PhysRevB.37.5571, moriarty1995first, mehta2006ab, sin2009ab, Grimvall_RevModPhys.84.945}  This instability is caused by an unstable transverse phonon mode at the bcc N-point zone boundary and is directly related to the mechanism of the martensitic hcp-bcc phase transformation~\cite{Wentzcovitch_PhysRevB.37.5571}.

\section{Conclusion}
\label{sec:conclusion}

We developed a general-purpose interatomic potential for Mg based on the atomic cluster expansion. The present ACE parametrization predicts materials properties with a similar accuracy as the DFT reference for a broad range of atomic environments. We demonstrated the broad applicability of the ACE model by analyzing properties of extended defects and evaluated the $P$-$T$ phase diagram of Mg over an extensive range of temperatures and pressures.

\section{Additional validation results}

\begin{figure}[hbt!]
\centering
\includegraphics[width=9cm,height=5cm]{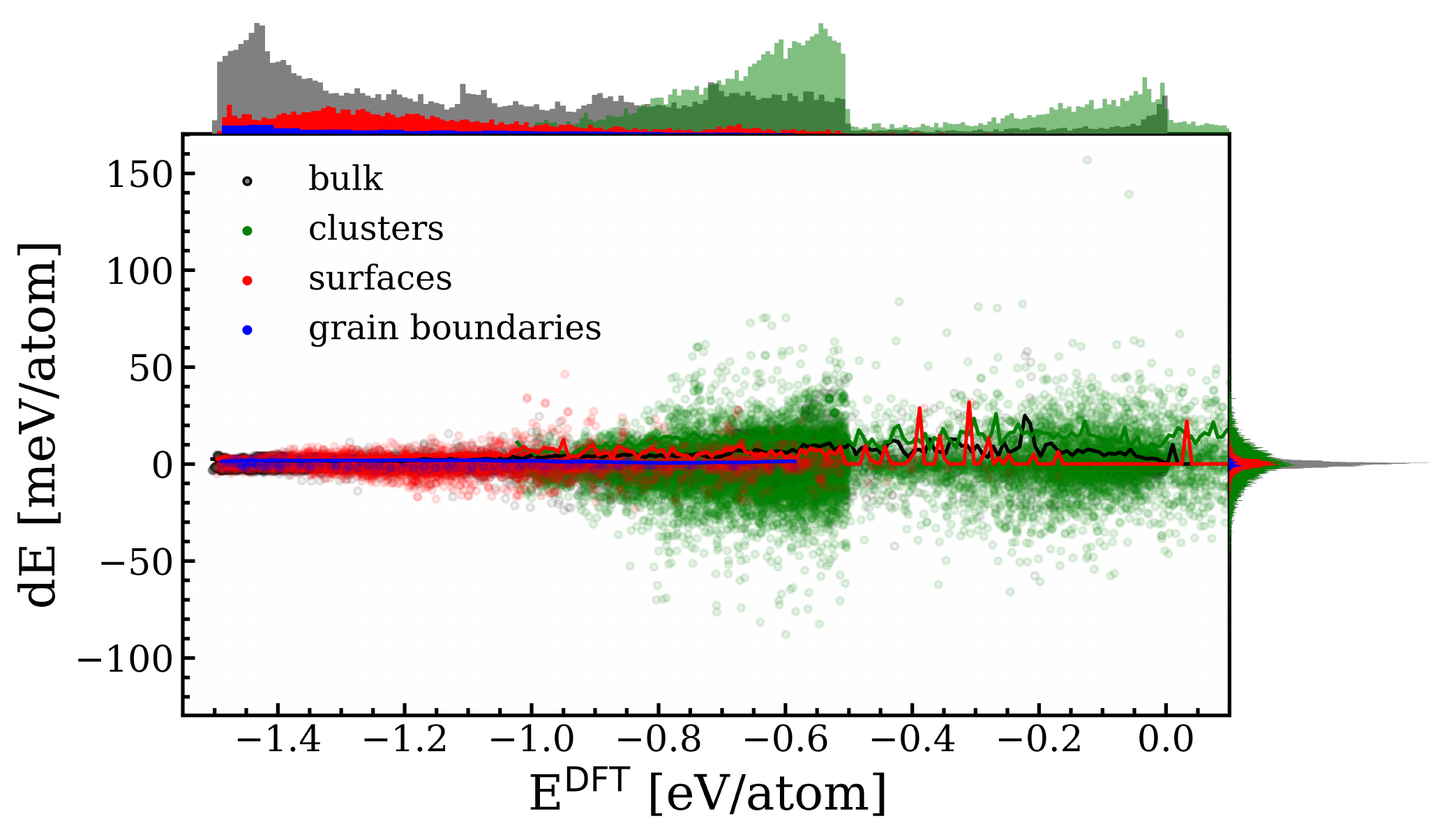}
\caption{Error in the energy per atom as a function of the reference DFT energy for the training dataset. The solid lines show the average
RMSE, the top panel shows the number of samples, and the right panel the overall error distribution.}
\label{train_e_de_dist}
\end{figure}

\begin{figure*}[hbt!]
\centering
\includegraphics
[width=15cm,height=12cm]{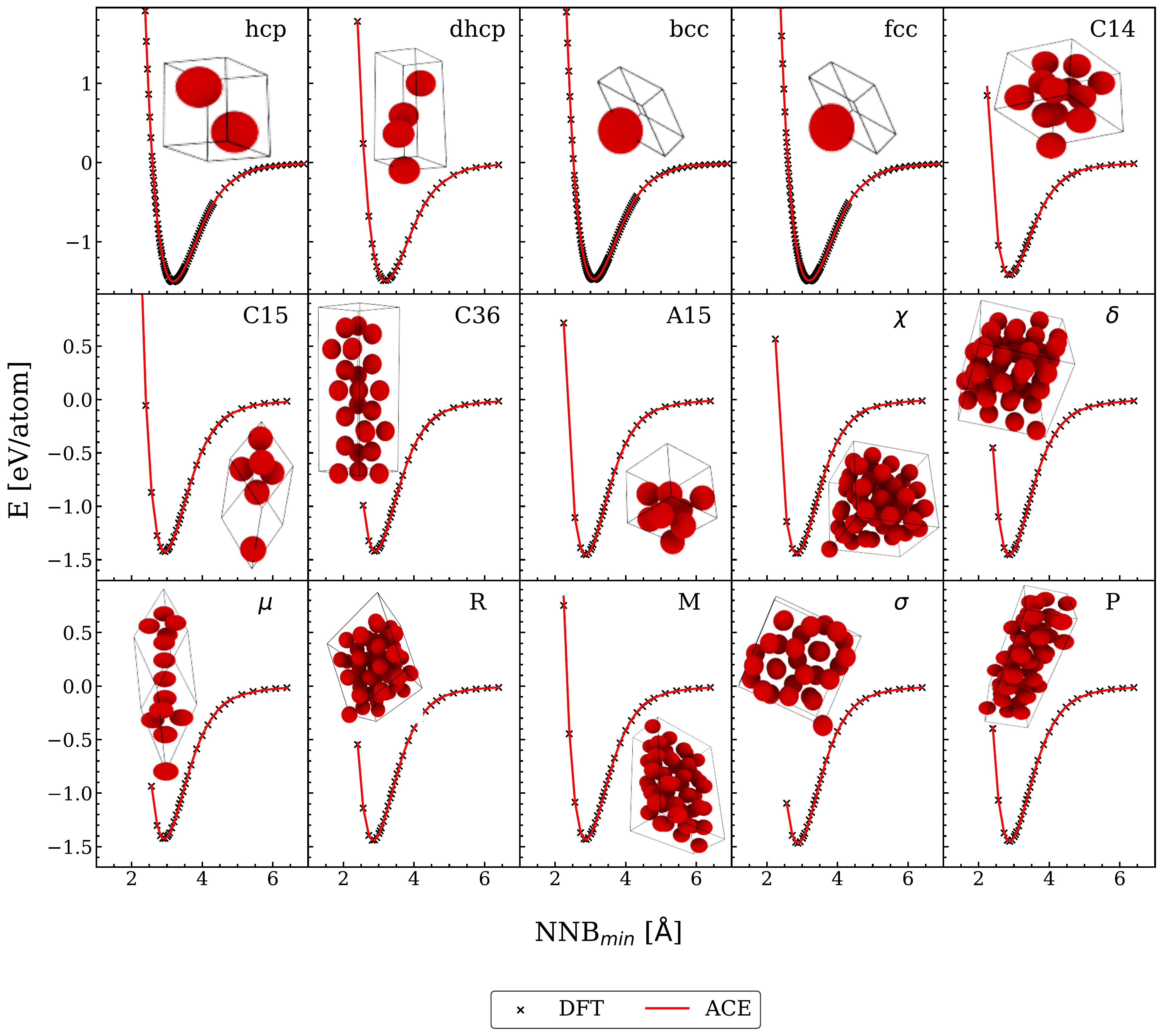}
\caption{Energy with respect to nearest neighbour distance for selected bulk structures.}
\label{e_nnb}
\end{figure*}

Our training dataset contains various structures including bulk, surfaces, grain boundaries, clusters, and various shaken and random structures.  In Fig.~\ref{e_nnb}, we show energy versus nearest neighbor distance (NNB$_{min}$) for additional bulk structures calculated using ACE and DFT. The results demonstrate a close agreement between ACE and the DFT reference. In addition, one can see that the ACE curves are smooth over the whole range of NNB$_{min}$ which is not straightforward to achieve. 

\begin{figure*}[hbt!]
\centering
\includegraphics[width=\textwidth]{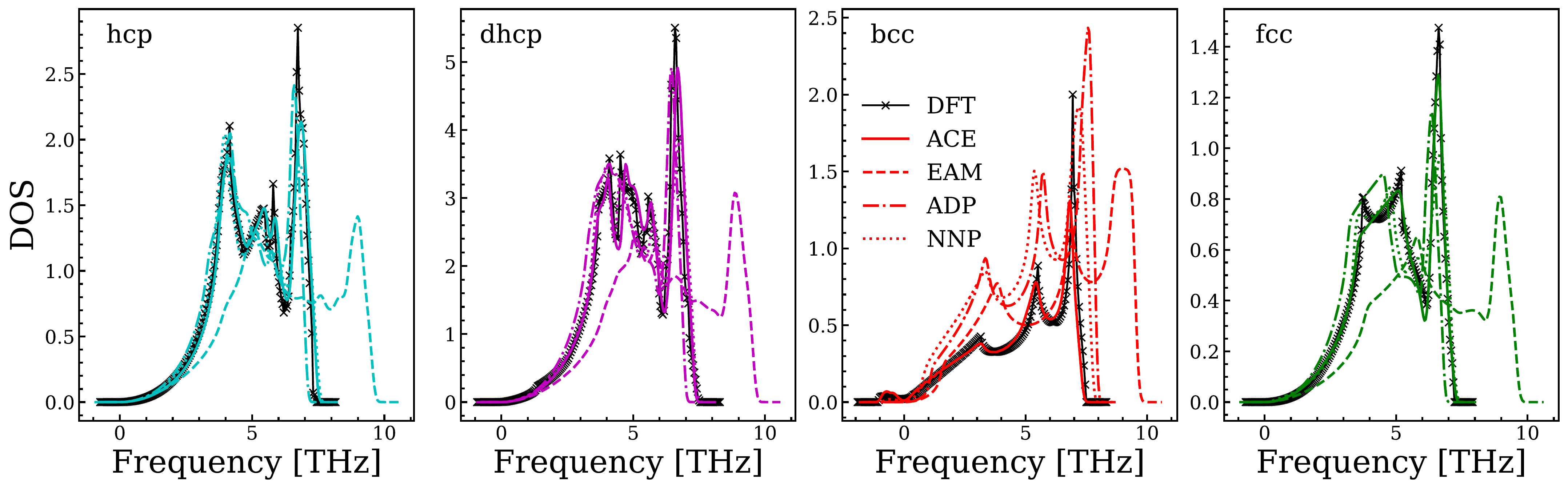}
\caption {Phonon DOS for hcp, dhcp, bcc, and fcc computed with DFT, ACE, EAM, ADP and NNP.}
\label{phon_ips}
\end{figure*}

\begin{figure*}[!tbh]
\centering
\includegraphics[width=15cm,height=10cm]{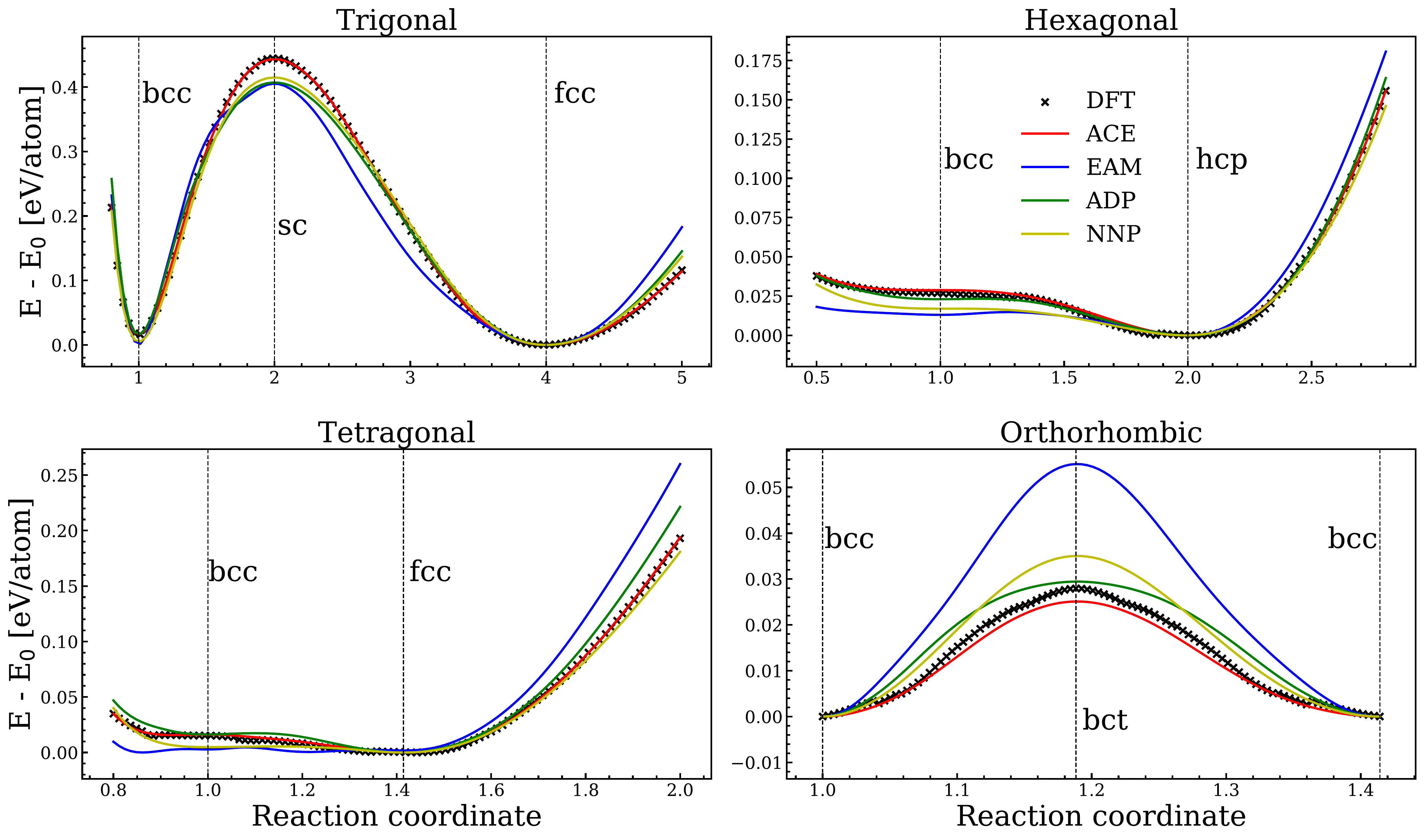}
\caption {Transformation paths for trigonal, hexagonal, tetragonal, and orthorhombic paths computed with DFT, ACE, EAM, ADP and NNP.}
\label{ips_tp}
\end{figure*}

In Fig.~\ref{train_e_de_dist}, we provide an additional quantification of the ACE errors and their distribution with respect to the DFT reference.

In Fig.~\ref{phon_ips}, we show a comparison of phonon DOS calculated using DFT, ACE and the considered interatomic potentials for four most relevant structures (hcp, dhcp, bcc and fcc). Only ACE correctly predicts imaginary phonon frequencies in the case of bcc in accordance with DFT, while EAM and ADP predicts the bcc structure to be dynamically stable.

The transformation paths for all considered interatomic potentials are compared in Fig.~\ref{ips_tp}.

\section{Free energy}

\begin{figure*}[hbt!]
\centering
\includegraphics[width=\textwidth]{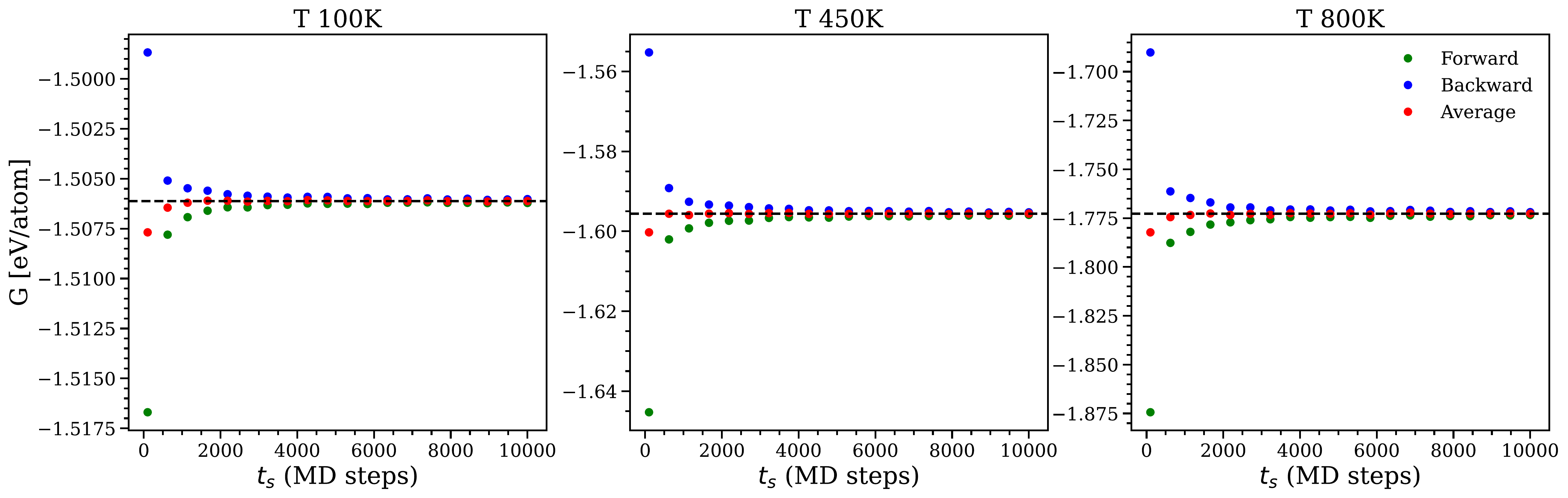}
\caption{FL NETI results for the forward direction, backward direction and their average for three different temperatures. The dashed line shows the value at the thermodynamic limit. The averaged path converges much faster than the forward and backward directions, showing the efficiency of the NETI scheme.}
\label{g_ts}
\end{figure*}

\begin{figure}[!tbh]
\centering
\includegraphics[width=9cm]{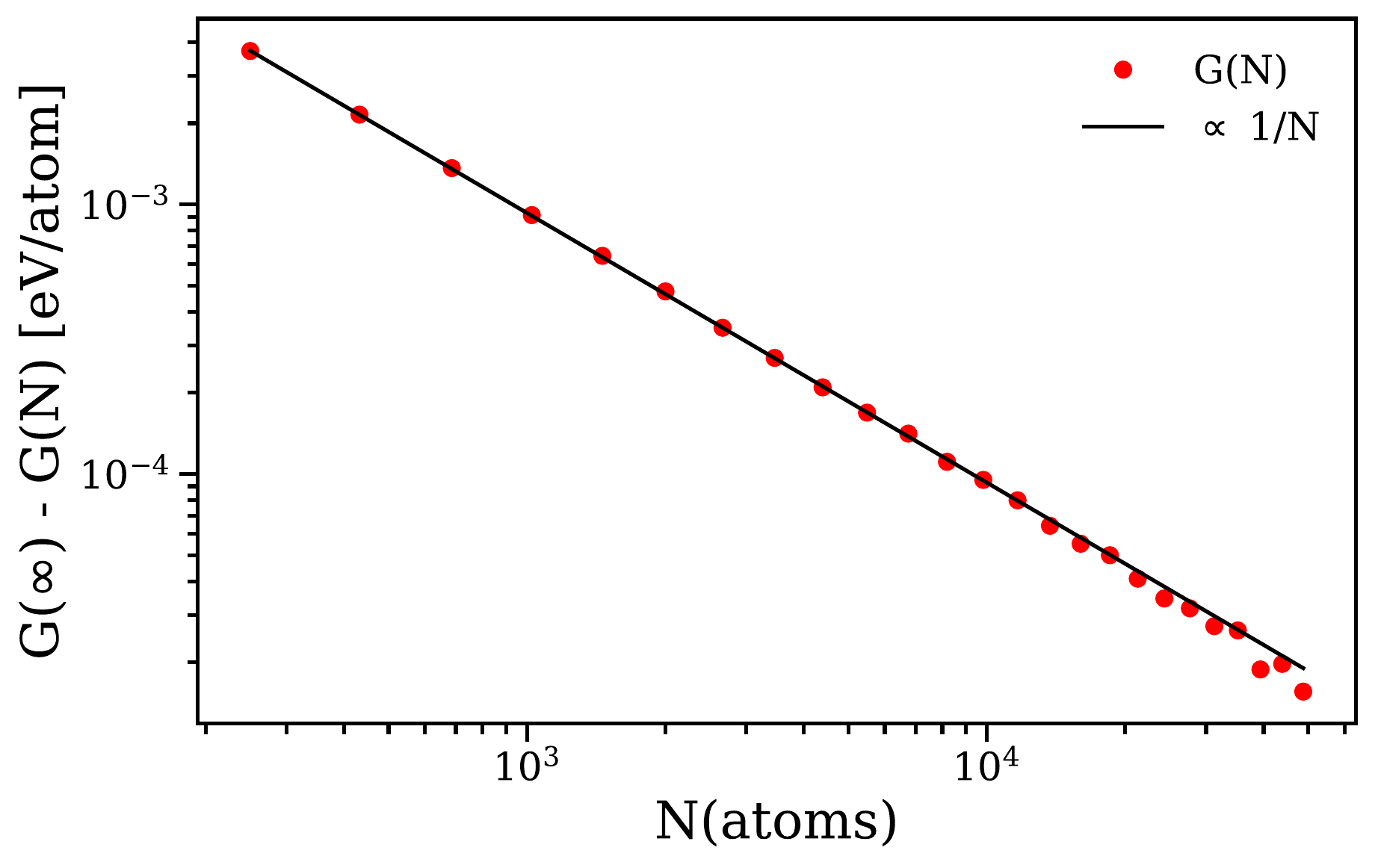}
\caption{We obtained an estimate for the Gibbs free energy in the thermodynamic limit, $G(\infty)$, by making an asymptotic analysis
which has shown that the free energy, $G(N)$, converges with the leading term 1/N.}
\label{g_inf}
\end{figure}

\begin{figure}[!tbh]
\centering
\includegraphics[width=9cm]{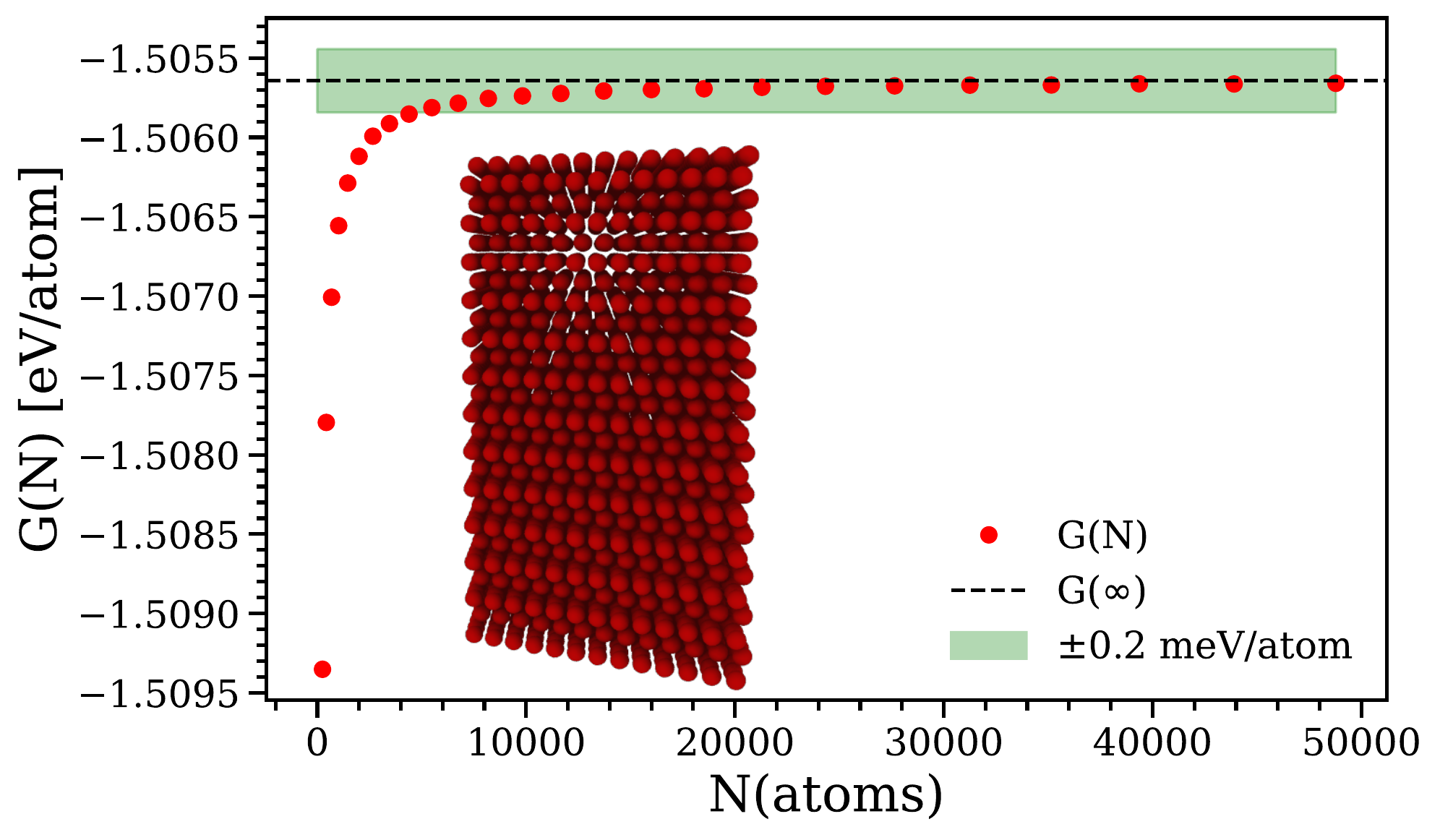}
\caption{Free-energy convergence with the system size for the hcp structure at 100 K and zero pressure. The chosen accuracy was within $\pm$ 0.2\, meV/atom of the free energy in the thermodynamic limit.}
\label{g_n}
\end{figure}

Here, we discuss the implementation and calculations of free energies using NETI. Figure~\ref{g_ts} shows the Gibbs free energies at zero pressure computed using NETI for three different temperatures of 100\,, 400\,, and 800\, K. The x-axis shows the switching time in terms of MD steps, where one MD step corresponds to 2\, fs. We employ the Frenkel-Ladd (FL) calculations in three directions, namely, forward, backward and average. The average direction indicates going one time forward and one time backward. Then, we average the results of both of them. The dashed line shows the free energy in the thermodynamic limit for each corresponding temperature. Here one sees that the average direction converges more rapidly than the forward and backward directions. This behavior explains how the heat of dissipation decreases differently with the different paths. Fig.~\ref{g_ts} shows that one obtains excellent free energy estimation within a few hundred MD steps. However, one gets the same accuracy of the average direction using either forward or backward direction within a few thousand MD steps. Irrespective of the simulation temperature, we see a typical behavior of free energy convergence. Naturally, the absolute values of the converged free energies are different for different temperatures. From another point of view, one can relate this to the rate of termination at which we carry out our process. Hence, one arrives at almost negligible heats of dissipation with the average direction at higher rates in comparison to either the forward or the backward direction, leading to a more computationally efficient way. In principle, this is due to two reasons. Firstly, we terminate the process at a relatively long switching time. Secondly, we average the forward and backward processes which have almost the same heats of dissipation even at shorter switching times. Thus, as we increase the switching time (i.e., decrease the rate at which we terminate our process.), we get smaller heats of dissipation from the forward and backward directions.

Another convergence test that we carried out is the behavior of the free energy
as a function of the system size. Fig.~\ref{g_n} shows a typical convergence behavior for the free energy as a function of the number of atoms or the system size. One sees that for a small number of atoms (approx. $<200$), deviations are high and cannot be ignored. However, for larger systems, the deviations are insignificant ($<1$ meV) and can be disregarded. In these analyses, we use the hcp lattice of Mg and we vary the dimensions N × N × N of the supercell where N = 5\,, 6\,, 7\,, ..., 29\,. This results in supercells containing from 250 to 48778\, atoms. From these results, one sees that attaining
a convergence of the free energy within a fraction of meV demands a few thousand
atoms, where we calculate the value of free energy at the thermodynamic limit as shown in Fig.~\ref{g_inf}.\\

\begin{figure}[!tbh]
\centering
\includegraphics[width=9cm]{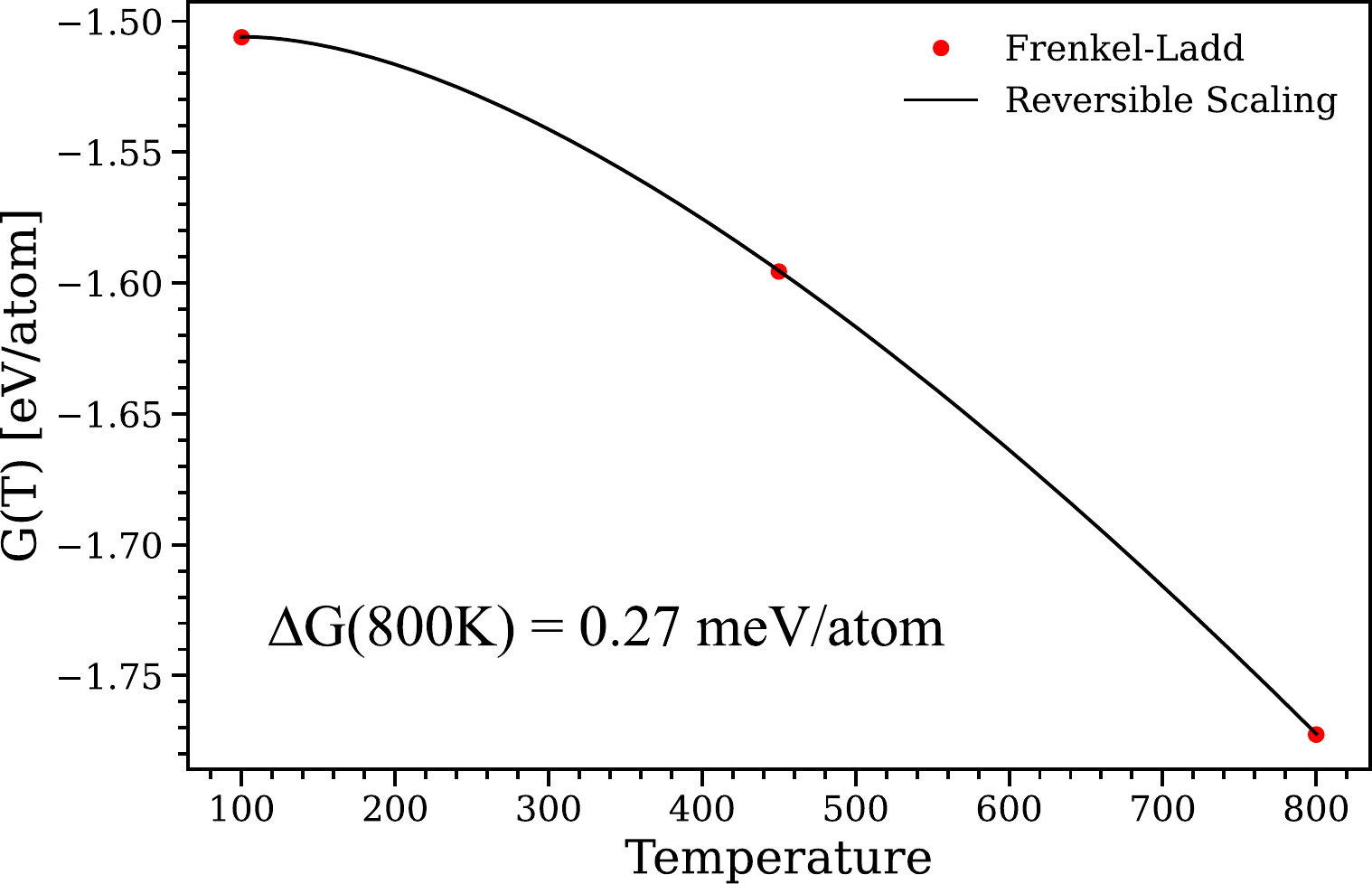}
\caption{Gibbs free energy of the hcp structure calculated using the ACE potential. The reference free energy from the FL thermodynamic path is at $T_{0}$ = 100 K and zero pressure. The FL calculations at
other temperatures were used to verify the agreement against the RS result at high temperatures.}
\label{rs}
\end{figure}

Figure~\ref{rs} shows the free energies from the FL and reversible scaling (RS) thermodynamic paths at zero pressure. We calculate the free energies using FL and RS thermodynamic paths in a temperature range from 100 to 800\, K (slightly below the melting point). Red points show single calculations of the free energies at temperatures of 100\, K, 450\, K, and 800\, K using FL thermodynamic path. The solid line shows the results of the free energies using the RS thermodynamic path with a reference point from FL free energy at 100\, K. In principle, RS calculations require a larger switching time to achieve similar accuracy as FL calculations. A typical FL switching time to reach convergence of a fraction of meV is around 5000\, MD steps, where an MD step corresponds to 2\, fs. A typical RS switching time is around 25000\, MD steps.\\

\begin{thebibliography}{79}%
\makeatletter
\providecommand \@ifxundefined [1]{%
 \@ifx{#1\undefined}
}%
\providecommand \@ifnum [1]{%
 \ifnum #1\expandafter \@firstoftwo
 \else \expandafter \@secondoftwo
 \fi
}%
\providecommand \@ifx [1]{%
 \ifx #1\expandafter \@firstoftwo
 \else \expandafter \@secondoftwo
 \fi
}%
\providecommand \natexlab [1]{#1}%
\providecommand \enquote  [1]{``#1''}%
\providecommand \bibnamefont  [1]{#1}%
\providecommand \bibfnamefont [1]{#1}%
\providecommand \citenamefont [1]{#1}%
\providecommand \href@noop [0]{\@secondoftwo}%
\providecommand \href [0]{\begingroup \@sanitize@url \@href}%
\providecommand \@href[1]{\@@startlink{#1}\@@href}%
\providecommand \@@href[1]{\endgroup#1\@@endlink}%
\providecommand \@sanitize@url [0]{\catcode `\\12\catcode `\$12\catcode
  `\&12\catcode `\#12\catcode `\^12\catcode `\_12\catcode `\%12\relax}%
\providecommand \@@startlink[1]{}%
\providecommand \@@endlink[0]{}%
\providecommand \url  [0]{\begingroup\@sanitize@url \@url }%
\providecommand \@url [1]{\endgroup\@href {#1}{\urlprefix }}%
\providecommand \urlprefix  [0]{URL }%
\providecommand \Eprint [0]{\href }%
\providecommand \doibase [0]{http://dx.doi.org/}%
\providecommand \selectlanguage [0]{\@gobble}%
\providecommand \bibinfo  [0]{\@secondoftwo}%
\providecommand \bibfield  [0]{\@secondoftwo}%
\providecommand \translation [1]{[#1]}%
\providecommand \BibitemOpen [0]{}%
\providecommand \bibitemStop [0]{}%
\providecommand \bibitemNoStop [0]{.\EOS\space}%
\providecommand \EOS [0]{\spacefactor3000\relax}%
\providecommand \BibitemShut  [1]{\csname bibitem#1\endcsname}%
\let\auto@bib@innerbib\@empty
\bibitem [{\citenamefont {Mordike}\ and\ \citenamefont
  {Ebert}(2001)}]{mordike2001magnesium}%
  \BibitemOpen
  \bibfield  {author} {\bibinfo {author} {\bibfnamefont {B.}~\bibnamefont
  {Mordike}}\ and\ \bibinfo {author} {\bibfnamefont {T.}~\bibnamefont
  {Ebert}},\ }\href@noop {} {\bibfield  {journal} {\bibinfo  {journal}
  {Materials Science and Engineering: A}\ }\textbf {\bibinfo {volume} {302}},\
  \bibinfo {pages} {37} (\bibinfo {year} {2001})}\BibitemShut {NoStop}%
\bibitem [{\citenamefont {Wu}\ and\ \citenamefont
  {Curtin}(2015)}]{wu2015origins}%
  \BibitemOpen
  \bibfield  {author} {\bibinfo {author} {\bibfnamefont {Z.}~\bibnamefont
  {Wu}}\ and\ \bibinfo {author} {\bibfnamefont {W.}~\bibnamefont {Curtin}},\
  }\href@noop {} {\bibfield  {journal} {\bibinfo  {journal} {Nature}\ }\textbf
  {\bibinfo {volume} {526}},\ \bibinfo {pages} {62} (\bibinfo {year}
  {2015})}\BibitemShut {NoStop}%
\bibitem [{\citenamefont {Pettifor}(1995)}]{pettifor1995bonding}%
  \BibitemOpen
  \bibfield  {author} {\bibinfo {author} {\bibfnamefont {D.~G.}\ \bibnamefont
  {Pettifor}},\ }\href@noop {} {\emph {\bibinfo {title} {Bonding and structure
  of molecules and solids}}}\ (\bibinfo  {publisher} {Oxford university
  press},\ \bibinfo {year} {1995})\BibitemShut {NoStop}%
\bibitem [{\citenamefont {Grimvall}\ \emph {et~al.}(2012)\citenamefont
  {Grimvall}, \citenamefont {Magyari-K\"ope}, \citenamefont {Ozolin\v{s}},\
  and\ \citenamefont {Persson}}]{Grimvall_RevModPhys.84.945}%
  \BibitemOpen
  \bibfield  {author} {\bibinfo {author} {\bibfnamefont {G.}~\bibnamefont
  {Grimvall}}, \bibinfo {author} {\bibfnamefont {B.}~\bibnamefont
  {Magyari-K\"ope}}, \bibinfo {author} {\bibfnamefont {V.}~\bibnamefont
  {Ozolin\v{s}}}, \ and\ \bibinfo {author} {\bibfnamefont {K.~A.}\ \bibnamefont
  {Persson}},\ }\href {\doibase 10.1103/RevModPhys.84.945} {\bibfield
  {journal} {\bibinfo  {journal} {Rev. Mod. Phys.}\ }\textbf {\bibinfo {volume}
  {84}},\ \bibinfo {pages} {945} (\bibinfo {year} {2012})}\BibitemShut
  {NoStop}%
\bibitem [{\citenamefont {McMahan}\ and\ \citenamefont
  {Moriarty}(1983)}]{mcmahan1983structural}%
  \BibitemOpen
  \bibfield  {author} {\bibinfo {author} {\bibfnamefont {A.}~\bibnamefont
  {McMahan}}\ and\ \bibinfo {author} {\bibfnamefont {J.~A.}\ \bibnamefont
  {Moriarty}},\ }\href@noop {} {\bibfield  {journal} {\bibinfo  {journal}
  {Physical Review B}\ }\textbf {\bibinfo {volume} {27}},\ \bibinfo {pages}
  {3235} (\bibinfo {year} {1983})}\BibitemShut {NoStop}%
\bibitem [{\citenamefont {McMahan}(1986)}]{mcmahan1986pressure}%
  \BibitemOpen
  \bibfield  {author} {\bibinfo {author} {\bibfnamefont {A.}~\bibnamefont
  {McMahan}},\ }\href@noop {} {\bibfield  {journal} {\bibinfo  {journal}
  {Physica B+ C}\ }\textbf {\bibinfo {volume} {139}},\ \bibinfo {pages} {31}
  (\bibinfo {year} {1986})}\BibitemShut {NoStop}%
\bibitem [{\citenamefont {Moriarty}\ and\ \citenamefont
  {Althoff}(1995)}]{moriarty1995first}%
  \BibitemOpen
  \bibfield  {author} {\bibinfo {author} {\bibfnamefont {J.~A.}\ \bibnamefont
  {Moriarty}}\ and\ \bibinfo {author} {\bibfnamefont {J.}~\bibnamefont
  {Althoff}},\ }\href@noop {} {\bibfield  {journal} {\bibinfo  {journal}
  {Physical Review B}\ }\textbf {\bibinfo {volume} {51}},\ \bibinfo {pages}
  {5609} (\bibinfo {year} {1995})}\BibitemShut {NoStop}%
\bibitem [{\citenamefont {Mehta}\ \emph {et~al.}(2006)\citenamefont {Mehta},
  \citenamefont {Price},\ and\ \citenamefont {Alf{\`e}}}]{mehta2006ab}%
  \BibitemOpen
  \bibfield  {author} {\bibinfo {author} {\bibfnamefont {S.}~\bibnamefont
  {Mehta}}, \bibinfo {author} {\bibfnamefont {G.}~\bibnamefont {Price}}, \ and\
  \bibinfo {author} {\bibfnamefont {D.}~\bibnamefont {Alf{\`e}}},\ }\href@noop
  {} {\bibfield  {journal} {\bibinfo  {journal} {The Journal of chemical
  physics}\ }\textbf {\bibinfo {volume} {125}},\ \bibinfo {pages} {194507}
  (\bibinfo {year} {2006})}\BibitemShut {NoStop}%
\bibitem [{\citenamefont {Li}\ \emph {et~al.}(2010)\citenamefont {Li},
  \citenamefont {Gao}, \citenamefont {Wang},\ and\ \citenamefont
  {Ma}}]{li2010crystal}%
  \BibitemOpen
  \bibfield  {author} {\bibinfo {author} {\bibfnamefont {P.}~\bibnamefont
  {Li}}, \bibinfo {author} {\bibfnamefont {G.}~\bibnamefont {Gao}}, \bibinfo
  {author} {\bibfnamefont {Y.}~\bibnamefont {Wang}}, \ and\ \bibinfo {author}
  {\bibfnamefont {Y.}~\bibnamefont {Ma}},\ }\href@noop {} {\bibfield  {journal}
  {\bibinfo  {journal} {The Journal of Physical Chemistry C}\ }\textbf
  {\bibinfo {volume} {114}},\ \bibinfo {pages} {21745} (\bibinfo {year}
  {2010})}\BibitemShut {NoStop}%
\bibitem [{\citenamefont {Cui}\ \emph {et~al.}(2022)\citenamefont {Cui},
  \citenamefont {Xian}, \citenamefont {Liu}, \citenamefont {Tian},
  \citenamefont {Gao},\ and\ \citenamefont {Song}}]{Cui2022}%
  \BibitemOpen
  \bibfield  {author} {\bibinfo {author} {\bibfnamefont {C.}~\bibnamefont
  {Cui}}, \bibinfo {author} {\bibfnamefont {J.}~\bibnamefont {Xian}}, \bibinfo
  {author} {\bibfnamefont {H.}~\bibnamefont {Liu}}, \bibinfo {author}
  {\bibfnamefont {F.}~\bibnamefont {Tian}}, \bibinfo {author} {\bibfnamefont
  {X.}~\bibnamefont {Gao}}, \ and\ \bibinfo {author} {\bibfnamefont
  {H.}~\bibnamefont {Song}},\ }\href {\doibase 10.1063/5.0087764} {\bibfield
  {journal} {\bibinfo  {journal} {Journal of Applied Physics}\ }\textbf
  {\bibinfo {volume} {131}},\ \bibinfo {pages} {195901} (\bibinfo {year}
  {2022})},\ \Eprint {http://arxiv.org/abs/https://doi.org/10.1063/5.0087764}
  {https://doi.org/10.1063/5.0087764} \BibitemShut {NoStop}%
\bibitem [{\citenamefont {Olijnyk}\ and\ \citenamefont
  {Holzapfel}(1985)}]{olijnyk1985high}%
  \BibitemOpen
  \bibfield  {author} {\bibinfo {author} {\bibfnamefont {H.}~\bibnamefont
  {Olijnyk}}\ and\ \bibinfo {author} {\bibfnamefont {W.}~\bibnamefont
  {Holzapfel}},\ }\href@noop {} {\bibfield  {journal} {\bibinfo  {journal}
  {Physical Review B}\ }\textbf {\bibinfo {volume} {31}},\ \bibinfo {pages}
  {4682} (\bibinfo {year} {1985})}\BibitemShut {NoStop}%
\bibitem [{\citenamefont {Errandonea}\ \emph {et~al.}(2001)\citenamefont
  {Errandonea}, \citenamefont {Boehler},\ and\ \citenamefont
  {Ross}}]{errandonea2001melting}%
  \BibitemOpen
  \bibfield  {author} {\bibinfo {author} {\bibfnamefont {D.}~\bibnamefont
  {Errandonea}}, \bibinfo {author} {\bibfnamefont {R.}~\bibnamefont {Boehler}},
  \ and\ \bibinfo {author} {\bibfnamefont {M.}~\bibnamefont {Ross}},\
  }\href@noop {} {\bibfield  {journal} {\bibinfo  {journal} {Physical Review
  B}\ }\textbf {\bibinfo {volume} {65}},\ \bibinfo {pages} {012108} (\bibinfo
  {year} {2001})}\BibitemShut {NoStop}%
\bibitem [{\citenamefont {Errandonea}\ \emph {et~al.}(2003)\citenamefont
  {Errandonea}, \citenamefont {Meng}, \citenamefont {Häusermann},\ and\
  \citenamefont {Uchida}}]{Errandonea_2003}%
  \BibitemOpen
  \bibfield  {author} {\bibinfo {author} {\bibfnamefont {D.}~\bibnamefont
  {Errandonea}}, \bibinfo {author} {\bibfnamefont {Y.}~\bibnamefont {Meng}},
  \bibinfo {author} {\bibfnamefont {D.}~\bibnamefont {Häusermann}}, \ and\
  \bibinfo {author} {\bibfnamefont {T.}~\bibnamefont {Uchida}},\ }\href
  {\doibase 10.1088/0953-8984/15/8/312} {\bibfield  {journal} {\bibinfo
  {journal} {Journal of Physics: Condensed Matter}\ }\textbf {\bibinfo {volume}
  {15}},\ \bibinfo {pages} {1277} (\bibinfo {year} {2003})}\BibitemShut
  {NoStop}%
\bibitem [{\citenamefont {Stinton}\ \emph {et~al.}(2014)\citenamefont
  {Stinton}, \citenamefont {MacLeod}, \citenamefont {Cynn}, \citenamefont
  {Errandonea}, \citenamefont {Evans}, \citenamefont {Proctor}, \citenamefont
  {Meng},\ and\ \citenamefont {McMahon}}]{Stinton_PhysRevB.90.134105}%
  \BibitemOpen
  \bibfield  {author} {\bibinfo {author} {\bibfnamefont {G.~W.}\ \bibnamefont
  {Stinton}}, \bibinfo {author} {\bibfnamefont {S.~G.}\ \bibnamefont
  {MacLeod}}, \bibinfo {author} {\bibfnamefont {H.}~\bibnamefont {Cynn}},
  \bibinfo {author} {\bibfnamefont {D.}~\bibnamefont {Errandonea}}, \bibinfo
  {author} {\bibfnamefont {W.~J.}\ \bibnamefont {Evans}}, \bibinfo {author}
  {\bibfnamefont {J.~E.}\ \bibnamefont {Proctor}}, \bibinfo {author}
  {\bibfnamefont {Y.}~\bibnamefont {Meng}}, \ and\ \bibinfo {author}
  {\bibfnamefont {M.~I.}\ \bibnamefont {McMahon}},\ }\href {\doibase
  10.1103/PhysRevB.90.134105} {\bibfield  {journal} {\bibinfo  {journal} {Phys.
  Rev. B}\ }\textbf {\bibinfo {volume} {90}},\ \bibinfo {pages} {134105}
  (\bibinfo {year} {2014})}\BibitemShut {NoStop}%
\bibitem [{\citenamefont {Beason}\ \emph {et~al.}(2021)\citenamefont {Beason},
  \citenamefont {Jensen},\ and\ \citenamefont
  {Crockett}}]{Beason_PhysRevB.104.144106}%
  \BibitemOpen
  \bibfield  {author} {\bibinfo {author} {\bibfnamefont {M.~T.}\ \bibnamefont
  {Beason}}, \bibinfo {author} {\bibfnamefont {B.~J.}\ \bibnamefont {Jensen}},
  \ and\ \bibinfo {author} {\bibfnamefont {S.~D.}\ \bibnamefont {Crockett}},\
  }\href {\doibase 10.1103/PhysRevB.104.144106} {\bibfield  {journal} {\bibinfo
   {journal} {Phys. Rev. B}\ }\textbf {\bibinfo {volume} {104}},\ \bibinfo
  {pages} {144106} (\bibinfo {year} {2021})}\BibitemShut {NoStop}%
\bibitem [{\citenamefont {Liu}\ and\ \citenamefont
  {Adams}(1998)}]{liu1998grain}%
  \BibitemOpen
  \bibfield  {author} {\bibinfo {author} {\bibfnamefont {X.-Y.}\ \bibnamefont
  {Liu}}\ and\ \bibinfo {author} {\bibfnamefont {J.~B.}\ \bibnamefont
  {Adams}},\ }\href@noop {} {\bibfield  {journal} {\bibinfo  {journal} {Acta
  Materialia}\ }\textbf {\bibinfo {volume} {46}},\ \bibinfo {pages} {3467}
  (\bibinfo {year} {1998})}\BibitemShut {NoStop}%
\bibitem [{\citenamefont {Zhou}\ \emph {et~al.}(2004)\citenamefont {Zhou},
  \citenamefont {Johnson},\ and\ \citenamefont {Wadley}}]{zhou2004misfit}%
  \BibitemOpen
  \bibfield  {author} {\bibinfo {author} {\bibfnamefont {X.}~\bibnamefont
  {Zhou}}, \bibinfo {author} {\bibfnamefont {R.}~\bibnamefont {Johnson}}, \
  and\ \bibinfo {author} {\bibfnamefont {H.}~\bibnamefont {Wadley}},\
  }\href@noop {} {\bibfield  {journal} {\bibinfo  {journal} {Physical Review
  B}\ }\textbf {\bibinfo {volume} {69}},\ \bibinfo {pages} {144113} (\bibinfo
  {year} {2004})}\BibitemShut {NoStop}%
\bibitem [{\citenamefont {Sun}\ \emph {et~al.}(2006)\citenamefont {Sun},
  \citenamefont {Mendelev}, \citenamefont {Becker}, \citenamefont {Kudin},
  \citenamefont {Haxhimali}, \citenamefont {Asta}, \citenamefont {Hoyt},
  \citenamefont {Karma},\ and\ \citenamefont {Srolovitz}}]{sun2006crystal}%
  \BibitemOpen
  \bibfield  {author} {\bibinfo {author} {\bibfnamefont {D.}~\bibnamefont
  {Sun}}, \bibinfo {author} {\bibfnamefont {M.}~\bibnamefont {Mendelev}},
  \bibinfo {author} {\bibfnamefont {C.}~\bibnamefont {Becker}}, \bibinfo
  {author} {\bibfnamefont {K.}~\bibnamefont {Kudin}}, \bibinfo {author}
  {\bibfnamefont {T.}~\bibnamefont {Haxhimali}}, \bibinfo {author}
  {\bibfnamefont {M.}~\bibnamefont {Asta}}, \bibinfo {author} {\bibfnamefont
  {J.}~\bibnamefont {Hoyt}}, \bibinfo {author} {\bibfnamefont {A.}~\bibnamefont
  {Karma}}, \ and\ \bibinfo {author} {\bibfnamefont {D.~J.}\ \bibnamefont
  {Srolovitz}},\ }\href@noop {} {\bibfield  {journal} {\bibinfo  {journal}
  {Physical Review B}\ }\textbf {\bibinfo {volume} {73}},\ \bibinfo {pages}
  {024116} (\bibinfo {year} {2006})}\BibitemShut {NoStop}%
\bibitem [{\citenamefont {Wilson}\ and\ \citenamefont
  {Mendelev}(2016)}]{wilson2016unified}%
  \BibitemOpen
  \bibfield  {author} {\bibinfo {author} {\bibfnamefont {S.}~\bibnamefont
  {Wilson}}\ and\ \bibinfo {author} {\bibfnamefont {M.}~\bibnamefont
  {Mendelev}},\ }\href@noop {} {\bibfield  {journal} {\bibinfo  {journal} {The
  Journal of Chemical Physics}\ }\textbf {\bibinfo {volume} {144}},\ \bibinfo
  {pages} {144707} (\bibinfo {year} {2016})}\BibitemShut {NoStop}%
\bibitem [{\citenamefont {Pei}\ \emph {et~al.}(2018)\citenamefont {Pei},
  \citenamefont {Sheng}, \citenamefont {Zhang}, \citenamefont {Li},\ and\
  \citenamefont {Svendsen}}]{pei2018tunable}%
  \BibitemOpen
  \bibfield  {author} {\bibinfo {author} {\bibfnamefont {Z.}~\bibnamefont
  {Pei}}, \bibinfo {author} {\bibfnamefont {H.}~\bibnamefont {Sheng}}, \bibinfo
  {author} {\bibfnamefont {X.}~\bibnamefont {Zhang}}, \bibinfo {author}
  {\bibfnamefont {R.}~\bibnamefont {Li}}, \ and\ \bibinfo {author}
  {\bibfnamefont {B.}~\bibnamefont {Svendsen}},\ }\href@noop {} {\bibfield
  {journal} {\bibinfo  {journal} {Materials \& Design}\ }\textbf {\bibinfo
  {volume} {153}},\ \bibinfo {pages} {232} (\bibinfo {year}
  {2018})}\BibitemShut {NoStop}%
\bibitem [{\citenamefont {Kim}\ \emph {et~al.}(2009)\citenamefont {Kim},
  \citenamefont {Kim},\ and\ \citenamefont {Lee}}]{kim2009atomistic}%
  \BibitemOpen
  \bibfield  {author} {\bibinfo {author} {\bibfnamefont {Y.-M.}\ \bibnamefont
  {Kim}}, \bibinfo {author} {\bibfnamefont {N.~J.}\ \bibnamefont {Kim}}, \ and\
  \bibinfo {author} {\bibfnamefont {B.-J.}\ \bibnamefont {Lee}},\ }\href@noop
  {} {\bibfield  {journal} {\bibinfo  {journal} {Calphad}\ }\textbf {\bibinfo
  {volume} {33}},\ \bibinfo {pages} {650} (\bibinfo {year} {2009})}\BibitemShut
  {NoStop}%
\bibitem [{\citenamefont {Dickel}\ \emph {et~al.}(2018)\citenamefont {Dickel},
  \citenamefont {Baskes}, \citenamefont {Aslam},\ and\ \citenamefont
  {Barrett}}]{dickel2018new}%
  \BibitemOpen
  \bibfield  {author} {\bibinfo {author} {\bibfnamefont {D.~E.}\ \bibnamefont
  {Dickel}}, \bibinfo {author} {\bibfnamefont {M.~I.}\ \bibnamefont {Baskes}},
  \bibinfo {author} {\bibfnamefont {I.}~\bibnamefont {Aslam}}, \ and\ \bibinfo
  {author} {\bibfnamefont {C.~D.}\ \bibnamefont {Barrett}},\ }\href@noop {}
  {\bibfield  {journal} {\bibinfo  {journal} {Modelling and simulation in
  materials science and engineering}\ }\textbf {\bibinfo {volume} {26}},\
  \bibinfo {pages} {045010} (\bibinfo {year} {2018})}\BibitemShut {NoStop}%
\bibitem [{\citenamefont {Ahmad}\ \emph {et~al.}(2020)\citenamefont {Ahmad},
  \citenamefont {Wu},\ and\ \citenamefont {Curtin}}]{ahmad2020analysis}%
  \BibitemOpen
  \bibfield  {author} {\bibinfo {author} {\bibfnamefont {R.}~\bibnamefont
  {Ahmad}}, \bibinfo {author} {\bibfnamefont {Z.}~\bibnamefont {Wu}}, \ and\
  \bibinfo {author} {\bibfnamefont {W.}~\bibnamefont {Curtin}},\ }\href@noop {}
  {\bibfield  {journal} {\bibinfo  {journal} {Acta Materialia}\ }\textbf
  {\bibinfo {volume} {183}},\ \bibinfo {pages} {228} (\bibinfo {year}
  {2020})}\BibitemShut {NoStop}%
\bibitem [{\citenamefont {Smirnova}\ \emph {et~al.}(2018)\citenamefont
  {Smirnova}, \citenamefont {Starikov},\ and\ \citenamefont
  {Vlasova}}]{smirnova2018new}%
  \BibitemOpen
  \bibfield  {author} {\bibinfo {author} {\bibfnamefont {D.}~\bibnamefont
  {Smirnova}}, \bibinfo {author} {\bibfnamefont {S.}~\bibnamefont {Starikov}},
  \ and\ \bibinfo {author} {\bibfnamefont {A.}~\bibnamefont {Vlasova}},\
  }\href@noop {} {\bibfield  {journal} {\bibinfo  {journal} {Computational
  Materials Science}\ }\textbf {\bibinfo {volume} {154}},\ \bibinfo {pages}
  {295} (\bibinfo {year} {2018})}\BibitemShut {NoStop}%
\bibitem [{\citenamefont {Cleri}\ and\ \citenamefont
  {Rosato}(1993)}]{cleri1993tight}%
  \BibitemOpen
  \bibfield  {author} {\bibinfo {author} {\bibfnamefont {F.}~\bibnamefont
  {Cleri}}\ and\ \bibinfo {author} {\bibfnamefont {V.}~\bibnamefont {Rosato}},\
  }\href@noop {} {\bibfield  {journal} {\bibinfo  {journal} {Physical Review
  B}\ }\textbf {\bibinfo {volume} {48}},\ \bibinfo {pages} {22} (\bibinfo
  {year} {1993})}\BibitemShut {NoStop}%
\bibitem [{\citenamefont {Li}\ \emph {et~al.}(2015)\citenamefont {Li},
  \citenamefont {Qin}, \citenamefont {Fu},\ and\ \citenamefont
  {Zhao}}]{li2015gupta}%
  \BibitemOpen
  \bibfield  {author} {\bibinfo {author} {\bibfnamefont {X.}~\bibnamefont
  {Li}}, \bibinfo {author} {\bibfnamefont {Y.}~\bibnamefont {Qin}}, \bibinfo
  {author} {\bibfnamefont {J.}~\bibnamefont {Fu}}, \ and\ \bibinfo {author}
  {\bibfnamefont {J.}~\bibnamefont {Zhao}},\ }\href@noop {} {\bibfield
  {journal} {\bibinfo  {journal} {Computational Materials Science}\ }\textbf
  {\bibinfo {volume} {98}},\ \bibinfo {pages} {328} (\bibinfo {year}
  {2015})}\BibitemShut {NoStop}%
\bibitem [{\citenamefont {Stricker}\ \emph {et~al.}(2020)\citenamefont
  {Stricker}, \citenamefont {Yin}, \citenamefont {Mak},\ and\ \citenamefont
  {Curtin}}]{stricker2020machine}%
  \BibitemOpen
  \bibfield  {author} {\bibinfo {author} {\bibfnamefont {M.}~\bibnamefont
  {Stricker}}, \bibinfo {author} {\bibfnamefont {B.}~\bibnamefont {Yin}},
  \bibinfo {author} {\bibfnamefont {E.}~\bibnamefont {Mak}}, \ and\ \bibinfo
  {author} {\bibfnamefont {W.}~\bibnamefont {Curtin}},\ }\href@noop {}
  {\bibfield  {journal} {\bibinfo  {journal} {Physical Review Materials}\
  }\textbf {\bibinfo {volume} {4}},\ \bibinfo {pages} {103602} (\bibinfo {year}
  {2020})}\BibitemShut {NoStop}%
\bibitem [{\citenamefont {Dickel}\ \emph {et~al.}(2021)\citenamefont {Dickel},
  \citenamefont {Nitol},\ and\ \citenamefont {Barrett}}]{dickel2021lammps}%
  \BibitemOpen
  \bibfield  {author} {\bibinfo {author} {\bibfnamefont {D.}~\bibnamefont
  {Dickel}}, \bibinfo {author} {\bibfnamefont {M.}~\bibnamefont {Nitol}}, \
  and\ \bibinfo {author} {\bibfnamefont {C.}~\bibnamefont {Barrett}},\
  }\href@noop {} {\bibfield  {journal} {\bibinfo  {journal} {Computational
  Materials Science}\ }\textbf {\bibinfo {volume} {196}},\ \bibinfo {pages}
  {110481} (\bibinfo {year} {2021})}\BibitemShut {NoStop}%
\bibitem [{\citenamefont {Poul}\ \emph {et~al.}(2023)\citenamefont {Poul},
  \citenamefont {Huber}, \citenamefont {Bitzek},\ and\ \citenamefont
  {Neugebauer}}]{Poul23}%
  \BibitemOpen
  \bibfield  {author} {\bibinfo {author} {\bibfnamefont {M.}~\bibnamefont
  {Poul}}, \bibinfo {author} {\bibfnamefont {L.}~\bibnamefont {Huber}},
  \bibinfo {author} {\bibfnamefont {E.}~\bibnamefont {Bitzek}}, \ and\ \bibinfo
  {author} {\bibfnamefont {J.}~\bibnamefont {Neugebauer}},\ }\href {\doibase
  10.1103/PhysRevB.107.104103} {\bibfield  {journal} {\bibinfo  {journal}
  {Phys. Rev. B}\ }\textbf {\bibinfo {volume} {107}},\ \bibinfo {pages}
  {104103} (\bibinfo {year} {2023})}\BibitemShut {NoStop}%
\bibitem [{\citenamefont {Troncoso}\ and\ \citenamefont
  {Turlo}(2022)}]{troncoso2022evaluating}%
  \BibitemOpen
  \bibfield  {author} {\bibinfo {author} {\bibfnamefont {J.~F.}\ \bibnamefont
  {Troncoso}}\ and\ \bibinfo {author} {\bibfnamefont {V.}~\bibnamefont
  {Turlo}},\ }\href@noop {} {\bibfield  {journal} {\bibinfo  {journal}
  {Modelling and Simulation in Materials Science and Engineering}\ }\textbf
  {\bibinfo {volume} {30}},\ \bibinfo {pages} {045009} (\bibinfo {year}
  {2022})}\BibitemShut {NoStop}%
\bibitem [{\citenamefont {Drautz}(2019)}]{drautz2019atomic}%
  \BibitemOpen
  \bibfield  {author} {\bibinfo {author} {\bibfnamefont {R.}~\bibnamefont
  {Drautz}},\ }\href@noop {} {\bibfield  {journal} {\bibinfo  {journal}
  {Physical Review B}\ }\textbf {\bibinfo {volume} {99}},\ \bibinfo {pages}
  {014104} (\bibinfo {year} {2019})}\BibitemShut {NoStop}%
\bibitem [{\citenamefont {Blum}\ \emph {et~al.}(2009)\citenamefont {Blum},
  \citenamefont {Gehrke}, \citenamefont {Hanke}, \citenamefont {Havu},
  \citenamefont {Havu}, \citenamefont {Ren}, \citenamefont {Reuter},\ and\
  \citenamefont {Scheffler}}]{fhi-aims1}%
  \BibitemOpen
  \bibfield  {author} {\bibinfo {author} {\bibfnamefont {V.}~\bibnamefont
  {Blum}}, \bibinfo {author} {\bibfnamefont {R.}~\bibnamefont {Gehrke}},
  \bibinfo {author} {\bibfnamefont {F.}~\bibnamefont {Hanke}}, \bibinfo
  {author} {\bibfnamefont {P.}~\bibnamefont {Havu}}, \bibinfo {author}
  {\bibfnamefont {V.}~\bibnamefont {Havu}}, \bibinfo {author} {\bibfnamefont
  {X.}~\bibnamefont {Ren}}, \bibinfo {author} {\bibfnamefont {K.}~\bibnamefont
  {Reuter}}, \ and\ \bibinfo {author} {\bibfnamefont {M.}~\bibnamefont
  {Scheffler}},\ }\href@noop {} {\bibfield  {journal} {\bibinfo  {journal}
  {Computer Physics Communications}\ }\textbf {\bibinfo {volume} {180}},\
  \bibinfo {pages} {2175} (\bibinfo {year} {2009})}\BibitemShut {NoStop}%
\bibitem [{\citenamefont {Havu}\ \emph {et~al.}(2009)\citenamefont {Havu},
  \citenamefont {Blum}, \citenamefont {Havu},\ and\ \citenamefont
  {Scheffler}}]{fhi-aims2}%
  \BibitemOpen
  \bibfield  {author} {\bibinfo {author} {\bibfnamefont {V.}~\bibnamefont
  {Havu}}, \bibinfo {author} {\bibfnamefont {V.}~\bibnamefont {Blum}}, \bibinfo
  {author} {\bibfnamefont {P.}~\bibnamefont {Havu}}, \ and\ \bibinfo {author}
  {\bibfnamefont {M.}~\bibnamefont {Scheffler}},\ }\href@noop {} {\bibfield
  {journal} {\bibinfo  {journal} {Journal of Computational Physics}\ }\textbf
  {\bibinfo {volume} {228}},\ \bibinfo {pages} {8367} (\bibinfo {year}
  {2009})}\BibitemShut {NoStop}%
\bibitem [{\citenamefont {Perdew}\ \emph {et~al.}(1996)\citenamefont {Perdew},
  \citenamefont {Burke},\ and\ \citenamefont
  {Ernzerhof}}]{perdew1996generalized}%
  \BibitemOpen
  \bibfield  {author} {\bibinfo {author} {\bibfnamefont {J.~P.}\ \bibnamefont
  {Perdew}}, \bibinfo {author} {\bibfnamefont {K.}~\bibnamefont {Burke}}, \
  and\ \bibinfo {author} {\bibfnamefont {M.}~\bibnamefont {Ernzerhof}},\
  }\href@noop {} {\bibfield  {journal} {\bibinfo  {journal} {Physical review
  letters}\ }\textbf {\bibinfo {volume} {77}},\ \bibinfo {pages} {3865}
  (\bibinfo {year} {1996})}\BibitemShut {NoStop}%
\bibitem [{\citenamefont {Bochkarev}\ \emph
  {et~al.}(2022{\natexlab{a}})\citenamefont {Bochkarev}, \citenamefont
  {Lysogorskiy}, \citenamefont {Menon}, \citenamefont {Qamar}, \citenamefont
  {Mrovec},\ and\ \citenamefont {Drautz}}]{bochkarev2022efficient}%
  \BibitemOpen
  \bibfield  {author} {\bibinfo {author} {\bibfnamefont {A.}~\bibnamefont
  {Bochkarev}}, \bibinfo {author} {\bibfnamefont {Y.}~\bibnamefont
  {Lysogorskiy}}, \bibinfo {author} {\bibfnamefont {S.}~\bibnamefont {Menon}},
  \bibinfo {author} {\bibfnamefont {M.}~\bibnamefont {Qamar}}, \bibinfo
  {author} {\bibfnamefont {M.}~\bibnamefont {Mrovec}}, \ and\ \bibinfo {author}
  {\bibfnamefont {R.}~\bibnamefont {Drautz}},\ }\href@noop {} {\bibfield
  {journal} {\bibinfo  {journal} {Physical Review Materials}\ }\textbf
  {\bibinfo {volume} {6}},\ \bibinfo {pages} {013804} (\bibinfo {year}
  {2022}{\natexlab{a}})}\BibitemShut {NoStop}%
\bibitem [{\citenamefont {Lysogorskiy}\ \emph {et~al.}(2021)\citenamefont
  {Lysogorskiy}, \citenamefont {Oord}, \citenamefont {Bochkarev}, \citenamefont
  {Menon}, \citenamefont {Rinaldi}, \citenamefont {Hammerschmidt},
  \citenamefont {Mrovec}, \citenamefont {Thompson}, \citenamefont {Cs{\'a}nyi},
  \citenamefont {Ortner} \emph {et~al.}}]{lysogorskiy2021performant}%
  \BibitemOpen
  \bibfield  {author} {\bibinfo {author} {\bibfnamefont {Y.}~\bibnamefont
  {Lysogorskiy}}, \bibinfo {author} {\bibfnamefont {C.~v.~d.}\ \bibnamefont
  {Oord}}, \bibinfo {author} {\bibfnamefont {A.}~\bibnamefont {Bochkarev}},
  \bibinfo {author} {\bibfnamefont {S.}~\bibnamefont {Menon}}, \bibinfo
  {author} {\bibfnamefont {M.}~\bibnamefont {Rinaldi}}, \bibinfo {author}
  {\bibfnamefont {T.}~\bibnamefont {Hammerschmidt}}, \bibinfo {author}
  {\bibfnamefont {M.}~\bibnamefont {Mrovec}}, \bibinfo {author} {\bibfnamefont
  {A.}~\bibnamefont {Thompson}}, \bibinfo {author} {\bibfnamefont
  {G.}~\bibnamefont {Cs{\'a}nyi}}, \bibinfo {author} {\bibfnamefont
  {C.}~\bibnamefont {Ortner}},  \emph {et~al.},\ }\href@noop {} {\bibfield
  {journal} {\bibinfo  {journal} {npj Computational Materials}\ }\textbf
  {\bibinfo {volume} {7}},\ \bibinfo {pages} {1} (\bibinfo {year}
  {2021})}\BibitemShut {NoStop}%
\bibitem [{\citenamefont {Qamar}\ \emph {et~al.}(2022)\citenamefont {Qamar},
  \citenamefont {Mrovec}, \citenamefont {Lysogorskiy}, \citenamefont
  {Bochkarev},\ and\ \citenamefont {Drautz}}]{qamar2022atomic}%
  \BibitemOpen
  \bibfield  {author} {\bibinfo {author} {\bibfnamefont {M.}~\bibnamefont
  {Qamar}}, \bibinfo {author} {\bibfnamefont {M.}~\bibnamefont {Mrovec}},
  \bibinfo {author} {\bibfnamefont {Y.}~\bibnamefont {Lysogorskiy}}, \bibinfo
  {author} {\bibfnamefont {A.}~\bibnamefont {Bochkarev}}, \ and\ \bibinfo
  {author} {\bibfnamefont {R.}~\bibnamefont {Drautz}},\ }\href@noop {}
  {\bibfield  {journal} {\bibinfo  {journal} {arXiv preprint arXiv:2210.09161}\
  } (\bibinfo {year} {2022})}\BibitemShut {NoStop}%
\bibitem [{\citenamefont {Bochkarev}\ \emph
  {et~al.}(2022{\natexlab{b}})\citenamefont {Bochkarev}, \citenamefont
  {Lysogorskiy}, \citenamefont {Ortner}, \citenamefont {Cs{\'a}nyi},\ and\
  \citenamefont {Drautz}}]{bochkarev2022multilayer}%
  \BibitemOpen
  \bibfield  {author} {\bibinfo {author} {\bibfnamefont {A.}~\bibnamefont
  {Bochkarev}}, \bibinfo {author} {\bibfnamefont {Y.}~\bibnamefont
  {Lysogorskiy}}, \bibinfo {author} {\bibfnamefont {C.}~\bibnamefont {Ortner}},
  \bibinfo {author} {\bibfnamefont {G.}~\bibnamefont {Cs{\'a}nyi}}, \ and\
  \bibinfo {author} {\bibfnamefont {R.}~\bibnamefont {Drautz}},\ }\href@noop {}
  {\bibfield  {journal} {\bibinfo  {journal} {arXiv preprint arXiv:2205.08177}\
  } (\bibinfo {year} {2022}{\natexlab{b}})}\BibitemShut {NoStop}%
\bibitem [{\citenamefont {Dusson}\ \emph {et~al.}(2022)\citenamefont {Dusson},
  \citenamefont {Bachmayr}, \citenamefont {Cs\'{a}nyi}, \citenamefont {Drautz},
  \citenamefont {Etter}, \citenamefont {{van der Oord}},\ and\ \citenamefont
  {Ortner}}]{Dusson22}%
  \BibitemOpen
  \bibfield  {author} {\bibinfo {author} {\bibfnamefont {G.}~\bibnamefont
  {Dusson}}, \bibinfo {author} {\bibfnamefont {M.}~\bibnamefont {Bachmayr}},
  \bibinfo {author} {\bibfnamefont {G.}~\bibnamefont {Cs\'{a}nyi}}, \bibinfo
  {author} {\bibfnamefont {R.}~\bibnamefont {Drautz}}, \bibinfo {author}
  {\bibfnamefont {S.}~\bibnamefont {Etter}}, \bibinfo {author} {\bibfnamefont
  {C.}~\bibnamefont {{van der Oord}}}, \ and\ \bibinfo {author} {\bibfnamefont
  {C.}~\bibnamefont {Ortner}},\ }\href {\doibase
  https://doi.org/10.1016/j.jcp.2022.110946} {\bibfield  {journal} {\bibinfo
  {journal} {Journal of Computational Physics}\ }\textbf {\bibinfo {volume}
  {454}},\ \bibinfo {pages} {110946} (\bibinfo {year} {2022})}\BibitemShut
  {NoStop}%
\bibitem [{\citenamefont {Drautz}(2020)}]{drautz2020atomic}%
  \BibitemOpen
  \bibfield  {author} {\bibinfo {author} {\bibfnamefont {R.}~\bibnamefont
  {Drautz}},\ }\href {\doibase 10.1103/PhysRevB.102.024104} {\bibfield
  {journal} {\bibinfo  {journal} {Phys. Rev. B}\ }\textbf {\bibinfo {volume}
  {102}},\ \bibinfo {pages} {024104} (\bibinfo {year} {2020})}\BibitemShut
  {NoStop}%
\bibitem [{\citenamefont {Thompson}\ \emph {et~al.}(2022)\citenamefont
  {Thompson}, \citenamefont {Aktulga}, \citenamefont {Berger}, \citenamefont
  {Bolintineanu}, \citenamefont {Brown}, \citenamefont {Crozier}, \citenamefont
  {in~'t Veld}, \citenamefont {Kohlmeyer}, \citenamefont {Moore}, \citenamefont
  {Nguyen}, \citenamefont {Shan}, \citenamefont {Stevens}, \citenamefont
  {Tranchida}, \citenamefont {Trott},\ and\ \citenamefont {Plimpton}}]{LAMMPS}%
  \BibitemOpen
  \bibfield  {author} {\bibinfo {author} {\bibfnamefont {A.~P.}\ \bibnamefont
  {Thompson}}, \bibinfo {author} {\bibfnamefont {H.~M.}\ \bibnamefont
  {Aktulga}}, \bibinfo {author} {\bibfnamefont {R.}~\bibnamefont {Berger}},
  \bibinfo {author} {\bibfnamefont {D.~S.}\ \bibnamefont {Bolintineanu}},
  \bibinfo {author} {\bibfnamefont {W.~M.}\ \bibnamefont {Brown}}, \bibinfo
  {author} {\bibfnamefont {P.~S.}\ \bibnamefont {Crozier}}, \bibinfo {author}
  {\bibfnamefont {P.~J.}\ \bibnamefont {in~'t Veld}}, \bibinfo {author}
  {\bibfnamefont {A.}~\bibnamefont {Kohlmeyer}}, \bibinfo {author}
  {\bibfnamefont {S.~G.}\ \bibnamefont {Moore}}, \bibinfo {author}
  {\bibfnamefont {T.~D.}\ \bibnamefont {Nguyen}}, \bibinfo {author}
  {\bibfnamefont {R.}~\bibnamefont {Shan}}, \bibinfo {author} {\bibfnamefont
  {M.~J.}\ \bibnamefont {Stevens}}, \bibinfo {author} {\bibfnamefont
  {J.}~\bibnamefont {Tranchida}}, \bibinfo {author} {\bibfnamefont
  {C.}~\bibnamefont {Trott}}, \ and\ \bibinfo {author} {\bibfnamefont {S.~J.}\
  \bibnamefont {Plimpton}},\ }\href {\doibase 10.1016/j.cpc.2021.108171}
  {\bibfield  {journal} {\bibinfo  {journal} {Comp. Phys. Comm.}\ }\textbf
  {\bibinfo {volume} {271}},\ \bibinfo {pages} {108171} (\bibinfo {year}
  {2022})}\BibitemShut {NoStop}%
\bibitem [{\citenamefont {Larsen}\ \emph {et~al.}(2017)\citenamefont {Larsen},
  \citenamefont {Mortensen}, \citenamefont {Blomqvist}, \citenamefont
  {Castelli}, \citenamefont {Christensen}, \citenamefont {Du{\l}ak},
  \citenamefont {Friis}, \citenamefont {Groves}, \citenamefont {Hammer},
  \citenamefont {Hargus} \emph {et~al.}}]{larsen2017atomic}%
  \BibitemOpen
  \bibfield  {author} {\bibinfo {author} {\bibfnamefont {A.~H.}\ \bibnamefont
  {Larsen}}, \bibinfo {author} {\bibfnamefont {J.~J.}\ \bibnamefont
  {Mortensen}}, \bibinfo {author} {\bibfnamefont {J.}~\bibnamefont
  {Blomqvist}}, \bibinfo {author} {\bibfnamefont {I.~E.}\ \bibnamefont
  {Castelli}}, \bibinfo {author} {\bibfnamefont {R.}~\bibnamefont
  {Christensen}}, \bibinfo {author} {\bibfnamefont {M.}~\bibnamefont
  {Du{\l}ak}}, \bibinfo {author} {\bibfnamefont {J.}~\bibnamefont {Friis}},
  \bibinfo {author} {\bibfnamefont {M.~N.}\ \bibnamefont {Groves}}, \bibinfo
  {author} {\bibfnamefont {B.}~\bibnamefont {Hammer}}, \bibinfo {author}
  {\bibfnamefont {C.}~\bibnamefont {Hargus}},  \emph {et~al.},\ }\href@noop {}
  {\bibfield  {journal} {\bibinfo  {journal} {Journal of Physics: Condensed
  Matter}\ }\textbf {\bibinfo {volume} {29}},\ \bibinfo {pages} {273002}
  (\bibinfo {year} {2017})}\BibitemShut {NoStop}%
\bibitem [{\citenamefont {Menon}\ \emph {et~al.}(2021)\citenamefont {Menon},
  \citenamefont {Lysogorskiy}, \citenamefont {Rogal},\ and\ \citenamefont
  {Drautz}}]{menon2021automated}%
  \BibitemOpen
  \bibfield  {author} {\bibinfo {author} {\bibfnamefont {S.}~\bibnamefont
  {Menon}}, \bibinfo {author} {\bibfnamefont {Y.}~\bibnamefont {Lysogorskiy}},
  \bibinfo {author} {\bibfnamefont {J.}~\bibnamefont {Rogal}}, \ and\ \bibinfo
  {author} {\bibfnamefont {R.}~\bibnamefont {Drautz}},\ }\href@noop {}
  {\bibfield  {journal} {\bibinfo  {journal} {Physical Review Materials}\
  }\textbf {\bibinfo {volume} {5}},\ \bibinfo {pages} {103801} (\bibinfo {year}
  {2021})}\BibitemShut {NoStop}%
\bibitem [{\citenamefont {Frenkel}\ and\ \citenamefont
  {Ladd}(1984)}]{frenkel1984new}%
  \BibitemOpen
  \bibfield  {author} {\bibinfo {author} {\bibfnamefont {D.}~\bibnamefont
  {Frenkel}}\ and\ \bibinfo {author} {\bibfnamefont {A.~J.}\ \bibnamefont
  {Ladd}},\ }\href@noop {} {\bibfield  {journal} {\bibinfo  {journal} {The
  Journal of chemical physics}\ }\textbf {\bibinfo {volume} {81}},\ \bibinfo
  {pages} {3188} (\bibinfo {year} {1984})}\BibitemShut {NoStop}%
\bibitem [{\citenamefont {de~Koning}\ \emph {et~al.}(1999)\citenamefont
  {de~Koning}, \citenamefont {Antonelli},\ and\ \citenamefont
  {Yip}}]{de1999optimized}%
  \BibitemOpen
  \bibfield  {author} {\bibinfo {author} {\bibfnamefont {M.}~\bibnamefont
  {de~Koning}}, \bibinfo {author} {\bibfnamefont {A.}~\bibnamefont
  {Antonelli}}, \ and\ \bibinfo {author} {\bibfnamefont {S.}~\bibnamefont
  {Yip}},\ }\href@noop {} {\bibfield  {journal} {\bibinfo  {journal} {Physical
  review letters}\ }\textbf {\bibinfo {volume} {83}},\ \bibinfo {pages} {3973}
  (\bibinfo {year} {1999})}\BibitemShut {NoStop}%
\bibitem [{\citenamefont {de~Koning}\ \emph {et~al.}(2001)\citenamefont
  {de~Koning}, \citenamefont {Antonelli},\ and\ \citenamefont
  {Yip}}]{de2001single}%
  \BibitemOpen
  \bibfield  {author} {\bibinfo {author} {\bibfnamefont {M.}~\bibnamefont
  {de~Koning}}, \bibinfo {author} {\bibfnamefont {A.}~\bibnamefont
  {Antonelli}}, \ and\ \bibinfo {author} {\bibfnamefont {S.}~\bibnamefont
  {Yip}},\ }\href@noop {} {\bibfield  {journal} {\bibinfo  {journal} {The
  Journal of Chemical Physics}\ }\textbf {\bibinfo {volume} {115}},\ \bibinfo
  {pages} {11025} (\bibinfo {year} {2001})}\BibitemShut {NoStop}%
\bibitem [{\citenamefont {Ibrahim}\ \emph {et~al.}()\citenamefont {Ibrahim},
  \citenamefont {Lysogorskiy}, \citenamefont {Mrovec},\ and\ \citenamefont
  {Drautz}}]{suppl}%
  \BibitemOpen
  \bibfield  {author} {\bibinfo {author} {\bibfnamefont {E.}~\bibnamefont
  {Ibrahim}}, \bibinfo {author} {\bibfnamefont {Y.}~\bibnamefont
  {Lysogorskiy}}, \bibinfo {author} {\bibfnamefont {M.}~\bibnamefont {Mrovec}},
  \ and\ \bibinfo {author} {\bibfnamefont {R.}~\bibnamefont {Drautz}},\
  }\href@noop {} {\ }\BibitemShut {NoStop}%
\bibitem [{\citenamefont {Janssen}\ \emph {et~al.}(2019)\citenamefont
  {Janssen}, \citenamefont {Surendralal}, \citenamefont {Lysogorskiy},
  \citenamefont {Todorova}, \citenamefont {Hickel}, \citenamefont {Drautz},\
  and\ \citenamefont {Neugebauer}}]{janssen2019pyiron}%
  \BibitemOpen
  \bibfield  {author} {\bibinfo {author} {\bibfnamefont {J.}~\bibnamefont
  {Janssen}}, \bibinfo {author} {\bibfnamefont {S.}~\bibnamefont
  {Surendralal}}, \bibinfo {author} {\bibfnamefont {Y.}~\bibnamefont
  {Lysogorskiy}}, \bibinfo {author} {\bibfnamefont {M.}~\bibnamefont
  {Todorova}}, \bibinfo {author} {\bibfnamefont {T.}~\bibnamefont {Hickel}},
  \bibinfo {author} {\bibfnamefont {R.}~\bibnamefont {Drautz}}, \ and\ \bibinfo
  {author} {\bibfnamefont {J.}~\bibnamefont {Neugebauer}},\ }\href@noop {}
  {\bibfield  {journal} {\bibinfo  {journal} {Computational Materials Science}\
  }\textbf {\bibinfo {volume} {163}},\ \bibinfo {pages} {24} (\bibinfo {year}
  {2019})}\BibitemShut {NoStop}%
\bibitem [{\citenamefont {Sin’ko}\ and\ \citenamefont
  {Smirnov}(2009)}]{sin2009ab}%
  \BibitemOpen
  \bibfield  {author} {\bibinfo {author} {\bibfnamefont {G.}~\bibnamefont
  {Sin’ko}}\ and\ \bibinfo {author} {\bibfnamefont {N.}~\bibnamefont
  {Smirnov}},\ }\href@noop {} {\bibfield  {journal} {\bibinfo  {journal}
  {Physical Review B}\ }\textbf {\bibinfo {volume} {80}},\ \bibinfo {pages}
  {104113} (\bibinfo {year} {2009})}\BibitemShut {NoStop}%
\bibitem [{\citenamefont {Slutsky}\ and\ \citenamefont
  {Garland}(1957)}]{slutsky1957elastic}%
  \BibitemOpen
  \bibfield  {author} {\bibinfo {author} {\bibfnamefont {L.~J.}\ \bibnamefont
  {Slutsky}}\ and\ \bibinfo {author} {\bibfnamefont {C.}~\bibnamefont
  {Garland}},\ }\href@noop {} {\bibfield  {journal} {\bibinfo  {journal}
  {Physical Review}\ }\textbf {\bibinfo {volume} {107}},\ \bibinfo {pages}
  {972} (\bibinfo {year} {1957})}\BibitemShut {NoStop}%
\bibitem [{\citenamefont {Togo}\ and\ \citenamefont
  {Tanaka}(2015)}]{togo2015first}%
  \BibitemOpen
  \bibfield  {author} {\bibinfo {author} {\bibfnamefont {A.}~\bibnamefont
  {Togo}}\ and\ \bibinfo {author} {\bibfnamefont {I.}~\bibnamefont {Tanaka}},\
  }\href@noop {} {\bibfield  {journal} {\bibinfo  {journal} {Scripta
  Materialia}\ }\textbf {\bibinfo {volume} {108}},\ \bibinfo {pages} {1}
  (\bibinfo {year} {2015})}\BibitemShut {NoStop}%
\bibitem [{\citenamefont {Kirklin}\ \emph {et~al.}(2015)\citenamefont
  {Kirklin}, \citenamefont {Saal}, \citenamefont {Meredig}, \citenamefont
  {Thompson}, \citenamefont {Doak}, \citenamefont {Aykol}, \citenamefont
  {R{\"u}hl},\ and\ \citenamefont {Wolverton}}]{kirklin2015open}%
  \BibitemOpen
  \bibfield  {author} {\bibinfo {author} {\bibfnamefont {S.}~\bibnamefont
  {Kirklin}}, \bibinfo {author} {\bibfnamefont {J.~E.}\ \bibnamefont {Saal}},
  \bibinfo {author} {\bibfnamefont {B.}~\bibnamefont {Meredig}}, \bibinfo
  {author} {\bibfnamefont {A.}~\bibnamefont {Thompson}}, \bibinfo {author}
  {\bibfnamefont {J.~W.}\ \bibnamefont {Doak}}, \bibinfo {author}
  {\bibfnamefont {M.}~\bibnamefont {Aykol}}, \bibinfo {author} {\bibfnamefont
  {S.}~\bibnamefont {R{\"u}hl}}, \ and\ \bibinfo {author} {\bibfnamefont
  {C.}~\bibnamefont {Wolverton}},\ }\href@noop {} {\bibfield  {journal}
  {\bibinfo  {journal} {npj Computational Materials}\ }\textbf {\bibinfo
  {volume} {1}},\ \bibinfo {pages} {1} (\bibinfo {year} {2015})}\BibitemShut
  {NoStop}%
\bibitem [{\citenamefont {Tran}\ \emph {et~al.}(2019)\citenamefont {Tran},
  \citenamefont {Li}, \citenamefont {Montoya}, \citenamefont {Winston},
  \citenamefont {Persson},\ and\ \citenamefont {Ong}}]{tran2019anisotropic}%
  \BibitemOpen
  \bibfield  {author} {\bibinfo {author} {\bibfnamefont {R.}~\bibnamefont
  {Tran}}, \bibinfo {author} {\bibfnamefont {X.-G.}\ \bibnamefont {Li}},
  \bibinfo {author} {\bibfnamefont {J.~H.}\ \bibnamefont {Montoya}}, \bibinfo
  {author} {\bibfnamefont {D.}~\bibnamefont {Winston}}, \bibinfo {author}
  {\bibfnamefont {K.~A.}\ \bibnamefont {Persson}}, \ and\ \bibinfo {author}
  {\bibfnamefont {S.~P.}\ \bibnamefont {Ong}},\ }\href@noop {} {\bibfield
  {journal} {\bibinfo  {journal} {Surface Science}\ }\textbf {\bibinfo {volume}
  {687}},\ \bibinfo {pages} {48} (\bibinfo {year} {2019})}\BibitemShut
  {NoStop}%
\bibitem [{\citenamefont {Zheng}\ \emph {et~al.}(2020)\citenamefont {Zheng},
  \citenamefont {Li}, \citenamefont {Tran}, \citenamefont {Chen}, \citenamefont
  {Horton}, \citenamefont {Winston}, \citenamefont {Persson},\ and\
  \citenamefont {Ong}}]{zheng2020grain}%
  \BibitemOpen
  \bibfield  {author} {\bibinfo {author} {\bibfnamefont {H.}~\bibnamefont
  {Zheng}}, \bibinfo {author} {\bibfnamefont {X.-G.}\ \bibnamefont {Li}},
  \bibinfo {author} {\bibfnamefont {R.}~\bibnamefont {Tran}}, \bibinfo {author}
  {\bibfnamefont {C.}~\bibnamefont {Chen}}, \bibinfo {author} {\bibfnamefont
  {M.}~\bibnamefont {Horton}}, \bibinfo {author} {\bibfnamefont
  {D.}~\bibnamefont {Winston}}, \bibinfo {author} {\bibfnamefont {K.~A.}\
  \bibnamefont {Persson}}, \ and\ \bibinfo {author} {\bibfnamefont {S.~P.}\
  \bibnamefont {Ong}},\ }\href@noop {} {\bibfield  {journal} {\bibinfo
  {journal} {Acta Materialia}\ }\textbf {\bibinfo {volume} {186}},\ \bibinfo
  {pages} {40} (\bibinfo {year} {2020})}\BibitemShut {NoStop}%
\bibitem [{\citenamefont {Rohrer}(2001)}]{rohrer2001structure}%
  \BibitemOpen
  \bibfield  {author} {\bibinfo {author} {\bibfnamefont {G.~S.}\ \bibnamefont
  {Rohrer}},\ }\href@noop {} {\emph {\bibinfo {title} {Structure and bonding in
  crystalline materials}}}\ (\bibinfo  {publisher} {Cambridge University
  Press},\ \bibinfo {year} {2001})\BibitemShut {NoStop}%
\bibitem [{\citenamefont {Hutchinson}\ and\ \citenamefont
  {Barnett}(2010)}]{hutchinson2010effective}%
  \BibitemOpen
  \bibfield  {author} {\bibinfo {author} {\bibfnamefont {W.}~\bibnamefont
  {Hutchinson}}\ and\ \bibinfo {author} {\bibfnamefont {M.}~\bibnamefont
  {Barnett}},\ }\href@noop {} {\bibfield  {journal} {\bibinfo  {journal}
  {Scripta Materialia}\ }\textbf {\bibinfo {volume} {63}},\ \bibinfo {pages}
  {737} (\bibinfo {year} {2010})}\BibitemShut {NoStop}%
\bibitem [{\citenamefont {S{\'a}nchez-Mart{\'\i}n}\ \emph
  {et~al.}(2014)\citenamefont {S{\'a}nchez-Mart{\'\i}n}, \citenamefont
  {P{\'e}rez-Prado}, \citenamefont {Segurado}, \citenamefont {Bohlen},
  \citenamefont {Guti{\'e}rrez-Urrutia}, \citenamefont {Llorca},\ and\
  \citenamefont {Molina-Aldareguia}}]{sanchez2014measuring}%
  \BibitemOpen
  \bibfield  {author} {\bibinfo {author} {\bibfnamefont {R.}~\bibnamefont
  {S{\'a}nchez-Mart{\'\i}n}}, \bibinfo {author} {\bibfnamefont {M.~T.}\
  \bibnamefont {P{\'e}rez-Prado}}, \bibinfo {author} {\bibfnamefont
  {J.}~\bibnamefont {Segurado}}, \bibinfo {author} {\bibfnamefont
  {J.}~\bibnamefont {Bohlen}}, \bibinfo {author} {\bibfnamefont
  {I.}~\bibnamefont {Guti{\'e}rrez-Urrutia}}, \bibinfo {author} {\bibfnamefont
  {J.}~\bibnamefont {Llorca}}, \ and\ \bibinfo {author} {\bibfnamefont {J.~M.}\
  \bibnamefont {Molina-Aldareguia}},\ }\href@noop {} {\bibfield  {journal}
  {\bibinfo  {journal} {Acta Materialia}\ }\textbf {\bibinfo {volume} {71}},\
  \bibinfo {pages} {283} (\bibinfo {year} {2014})}\BibitemShut {NoStop}%
\bibitem [{\citenamefont {Fan}\ and\ \citenamefont
  {El-Awady}(2015)}]{fan2015towards}%
  \BibitemOpen
  \bibfield  {author} {\bibinfo {author} {\bibfnamefont {H.}~\bibnamefont
  {Fan}}\ and\ \bibinfo {author} {\bibfnamefont {J.~A.}\ \bibnamefont
  {El-Awady}},\ }\href@noop {} {\bibfield  {journal} {\bibinfo  {journal}
  {Materials Science and Engineering: A}\ }\textbf {\bibinfo {volume} {644}},\
  \bibinfo {pages} {318} (\bibinfo {year} {2015})}\BibitemShut {NoStop}%
\bibitem [{\citenamefont {Geng}\ \emph {et~al.}(2014)\citenamefont {Geng},
  \citenamefont {Chisholm}, \citenamefont {Mishra},\ and\ \citenamefont
  {Kumar}}]{geng2014structure}%
  \BibitemOpen
  \bibfield  {author} {\bibinfo {author} {\bibfnamefont {J.}~\bibnamefont
  {Geng}}, \bibinfo {author} {\bibfnamefont {M.~F.}\ \bibnamefont {Chisholm}},
  \bibinfo {author} {\bibfnamefont {R.}~\bibnamefont {Mishra}}, \ and\ \bibinfo
  {author} {\bibfnamefont {K.}~\bibnamefont {Kumar}},\ }\href@noop {}
  {\bibfield  {journal} {\bibinfo  {journal} {Philosophical Magazine Letters}\
  }\textbf {\bibinfo {volume} {94}},\ \bibinfo {pages} {377} (\bibinfo {year}
  {2014})}\BibitemShut {NoStop}%
\bibitem [{\citenamefont {Itakura}\ \emph {et~al.}(2016)\citenamefont
  {Itakura}, \citenamefont {Kaburaki}, \citenamefont {Yamaguchi},\ and\
  \citenamefont {Tsuru}}]{Itakura2016}%
  \BibitemOpen
  \bibfield  {author} {\bibinfo {author} {\bibfnamefont {M.}~\bibnamefont
  {Itakura}}, \bibinfo {author} {\bibfnamefont {H.}~\bibnamefont {Kaburaki}},
  \bibinfo {author} {\bibfnamefont {M.}~\bibnamefont {Yamaguchi}}, \ and\
  \bibinfo {author} {\bibfnamefont {T.}~\bibnamefont {Tsuru}},\ }\href
  {\doibase 10.1103/PhysRevLett.116.225501} {\bibfield  {journal} {\bibinfo
  {journal} {Phys. Rev. Lett.}\ }\textbf {\bibinfo {volume} {116}},\ \bibinfo
  {pages} {225501} (\bibinfo {year} {2016})}\BibitemShut {NoStop}%
\bibitem [{\citenamefont {Xie}\ \emph {et~al.}(2016)\citenamefont {Xie},
  \citenamefont {Alam}, \citenamefont {Caffee},\ and\ \citenamefont
  {Hemker}}]{xie2016pyramidal}%
  \BibitemOpen
  \bibfield  {author} {\bibinfo {author} {\bibfnamefont {K.~Y.}\ \bibnamefont
  {Xie}}, \bibinfo {author} {\bibfnamefont {Z.}~\bibnamefont {Alam}}, \bibinfo
  {author} {\bibfnamefont {A.}~\bibnamefont {Caffee}}, \ and\ \bibinfo {author}
  {\bibfnamefont {K.~J.}\ \bibnamefont {Hemker}},\ }\href@noop {} {\bibfield
  {journal} {\bibinfo  {journal} {Scripta Materialia}\ }\textbf {\bibinfo
  {volume} {112}},\ \bibinfo {pages} {75} (\bibinfo {year} {2016})}\BibitemShut
  {NoStop}%
\bibitem [{\citenamefont {Yin}\ \emph {et~al.}(2017)\citenamefont {Yin},
  \citenamefont {Wu},\ and\ \citenamefont {Curtin}}]{yin2017comprehensive}%
  \BibitemOpen
  \bibfield  {author} {\bibinfo {author} {\bibfnamefont {B.}~\bibnamefont
  {Yin}}, \bibinfo {author} {\bibfnamefont {Z.}~\bibnamefont {Wu}}, \ and\
  \bibinfo {author} {\bibfnamefont {W.}~\bibnamefont {Curtin}},\ }\href@noop {}
  {\bibfield  {journal} {\bibinfo  {journal} {Acta Materialia}\ }\textbf
  {\bibinfo {volume} {123}},\ \bibinfo {pages} {223} (\bibinfo {year}
  {2017})}\BibitemShut {NoStop}%
\bibitem [{\citenamefont {Vitek}(1968)}]{vitek1968intrinsic}%
  \BibitemOpen
  \bibfield  {author} {\bibinfo {author} {\bibfnamefont {V.}~\bibnamefont
  {Vitek}},\ }\href@noop {} {\bibfield  {journal} {\bibinfo  {journal}
  {Philosophical Magazine}\ }\textbf {\bibinfo {volume} {18}},\ \bibinfo
  {pages} {773} (\bibinfo {year} {1968})}\BibitemShut {NoStop}%
\bibitem [{\citenamefont {Chetty}\ and\ \citenamefont
  {Weinert}(1997)}]{chetty1997stacking}%
  \BibitemOpen
  \bibfield  {author} {\bibinfo {author} {\bibfnamefont {N.}~\bibnamefont
  {Chetty}}\ and\ \bibinfo {author} {\bibfnamefont {M.}~\bibnamefont
  {Weinert}},\ }\href@noop {} {\bibfield  {journal} {\bibinfo  {journal}
  {Physical Review B}\ }\textbf {\bibinfo {volume} {56}},\ \bibinfo {pages}
  {10844} (\bibinfo {year} {1997})}\BibitemShut {NoStop}%
\bibitem [{\citenamefont {Smith}(2007)}]{smith2007surface}%
  \BibitemOpen
  \bibfield  {author} {\bibinfo {author} {\bibfnamefont {A.~E.}\ \bibnamefont
  {Smith}},\ }\href@noop {} {\bibfield  {journal} {\bibinfo  {journal} {Surface
  Science}\ }\textbf {\bibinfo {volume} {601}},\ \bibinfo {pages} {5762}
  (\bibinfo {year} {2007})}\BibitemShut {NoStop}%
\bibitem [{\citenamefont {Shin}\ and\ \citenamefont
  {Carter}(2011)}]{shin2011orbital}%
  \BibitemOpen
  \bibfield  {author} {\bibinfo {author} {\bibfnamefont {I.}~\bibnamefont
  {Shin}}\ and\ \bibinfo {author} {\bibfnamefont {E.~A.}\ \bibnamefont
  {Carter}},\ }\href@noop {} {\bibfield  {journal} {\bibinfo  {journal}
  {Modelling and Simulation in Materials Science and Engineering}\ }\textbf
  {\bibinfo {volume} {20}},\ \bibinfo {pages} {015006} (\bibinfo {year}
  {2011})}\BibitemShut {NoStop}%
\bibitem [{\citenamefont {Wen}\ \emph {et~al.}(2009)\citenamefont {Wen},
  \citenamefont {Chen}, \citenamefont {Tong}, \citenamefont {Tang},
  \citenamefont {Peng},\ and\ \citenamefont {Ding}}]{wen2009systematic}%
  \BibitemOpen
  \bibfield  {author} {\bibinfo {author} {\bibfnamefont {L.}~\bibnamefont
  {Wen}}, \bibinfo {author} {\bibfnamefont {P.}~\bibnamefont {Chen}}, \bibinfo
  {author} {\bibfnamefont {Z.-F.}\ \bibnamefont {Tong}}, \bibinfo {author}
  {\bibfnamefont {B.-Y.}\ \bibnamefont {Tang}}, \bibinfo {author}
  {\bibfnamefont {L.-M.}\ \bibnamefont {Peng}}, \ and\ \bibinfo {author}
  {\bibfnamefont {W.-J.}\ \bibnamefont {Ding}},\ }\href@noop {} {\bibfield
  {journal} {\bibinfo  {journal} {The European Physical Journal B}\ }\textbf
  {\bibinfo {volume} {72}},\ \bibinfo {pages} {397} (\bibinfo {year}
  {2009})}\BibitemShut {NoStop}%
\bibitem [{\citenamefont {Wang}\ \emph {et~al.}(2010)\citenamefont {Wang},
  \citenamefont {Chen}, \citenamefont {Liu},\ and\ \citenamefont
  {Mathaudhu}}]{wang2010first}%
  \BibitemOpen
  \bibfield  {author} {\bibinfo {author} {\bibfnamefont {Y.}~\bibnamefont
  {Wang}}, \bibinfo {author} {\bibfnamefont {L.-Q.}\ \bibnamefont {Chen}},
  \bibinfo {author} {\bibfnamefont {Z.-K.}\ \bibnamefont {Liu}}, \ and\
  \bibinfo {author} {\bibfnamefont {S.}~\bibnamefont {Mathaudhu}},\ }\href@noop
  {} {\bibfield  {journal} {\bibinfo  {journal} {Scripta Materialia}\ }\textbf
  {\bibinfo {volume} {62}},\ \bibinfo {pages} {646} (\bibinfo {year}
  {2010})}\BibitemShut {NoStop}%
\bibitem [{\citenamefont {Zhang}\ \emph {et~al.}(2013)\citenamefont {Zhang},
  \citenamefont {Dou}, \citenamefont {Liu},\ and\ \citenamefont
  {Guo}}]{zhang2013first}%
  \BibitemOpen
  \bibfield  {author} {\bibinfo {author} {\bibfnamefont {J.}~\bibnamefont
  {Zhang}}, \bibinfo {author} {\bibfnamefont {Y.}~\bibnamefont {Dou}}, \bibinfo
  {author} {\bibfnamefont {G.}~\bibnamefont {Liu}}, \ and\ \bibinfo {author}
  {\bibfnamefont {Z.}~\bibnamefont {Guo}},\ }\href@noop {} {\bibfield
  {journal} {\bibinfo  {journal} {Computational materials science}\ }\textbf
  {\bibinfo {volume} {79}},\ \bibinfo {pages} {564} (\bibinfo {year}
  {2013})}\BibitemShut {NoStop}%
\bibitem [{\citenamefont {Smallman}\ and\ \citenamefont
  {Dobson}(1970)}]{smallman1970stacking}%
  \BibitemOpen
  \bibfield  {author} {\bibinfo {author} {\bibfnamefont {R.}~\bibnamefont
  {Smallman}}\ and\ \bibinfo {author} {\bibfnamefont {P.}~\bibnamefont
  {Dobson}},\ }\href@noop {} {\bibfield  {journal} {\bibinfo  {journal}
  {Metallurgical Transactions}\ }\textbf {\bibinfo {volume} {1}},\ \bibinfo
  {pages} {2383} (\bibinfo {year} {1970})}\BibitemShut {NoStop}%
\bibitem [{\citenamefont {Sastry}\ \emph {et~al.}(1969)\citenamefont {Sastry},
  \citenamefont {Prasad},\ and\ \citenamefont {Vasu}}]{sastry1969stacking}%
  \BibitemOpen
  \bibfield  {author} {\bibinfo {author} {\bibfnamefont {D.}~\bibnamefont
  {Sastry}}, \bibinfo {author} {\bibfnamefont {Y.}~\bibnamefont {Prasad}}, \
  and\ \bibinfo {author} {\bibfnamefont {K.}~\bibnamefont {Vasu}},\ }\href@noop
  {} {\bibfield  {journal} {\bibinfo  {journal} {Scripta Metallurgica}\
  }\textbf {\bibinfo {volume} {3}},\ \bibinfo {pages} {927} (\bibinfo {year}
  {1969})}\BibitemShut {NoStop}%
\bibitem [{\citenamefont {Couret}\ and\ \citenamefont
  {Caillard}(1985)}]{couret1985situ}%
  \BibitemOpen
  \bibfield  {author} {\bibinfo {author} {\bibfnamefont {A.}~\bibnamefont
  {Couret}}\ and\ \bibinfo {author} {\bibfnamefont {D.}~\bibnamefont
  {Caillard}},\ }\href@noop {} {\bibfield  {journal} {\bibinfo  {journal} {Acta
  Metallurgica}\ }\textbf {\bibinfo {volume} {33}},\ \bibinfo {pages} {1455}
  (\bibinfo {year} {1985})}\BibitemShut {NoStop}%
\bibitem [{\citenamefont {Yasi}\ \emph {et~al.}(2009)\citenamefont {Yasi},
  \citenamefont {Nogaret}, \citenamefont {Trinkle}, \citenamefont {Qi},
  \citenamefont {Hector},\ and\ \citenamefont {Curtin}}]{yasi2009basal}%
  \BibitemOpen
  \bibfield  {author} {\bibinfo {author} {\bibfnamefont {J.}~\bibnamefont
  {Yasi}}, \bibinfo {author} {\bibfnamefont {T.}~\bibnamefont {Nogaret}},
  \bibinfo {author} {\bibfnamefont {D.}~\bibnamefont {Trinkle}}, \bibinfo
  {author} {\bibfnamefont {Y.}~\bibnamefont {Qi}}, \bibinfo {author}
  {\bibfnamefont {L.}~\bibnamefont {Hector}}, \ and\ \bibinfo {author}
  {\bibfnamefont {W.}~\bibnamefont {Curtin}},\ }\href@noop {} {\bibfield
  {journal} {\bibinfo  {journal} {Modelling and Simulation in Materials Science
  and Engineering}\ }\textbf {\bibinfo {volume} {17}},\ \bibinfo {pages}
  {055012} (\bibinfo {year} {2009})}\BibitemShut {NoStop}%
\bibitem [{\citenamefont {Ghazisaeidi}\ \emph {et~al.}(2014)\citenamefont
  {Ghazisaeidi}, \citenamefont {Hector~Jr},\ and\ \citenamefont
  {Curtin}}]{ghazisaeidi2014first}%
  \BibitemOpen
  \bibfield  {author} {\bibinfo {author} {\bibfnamefont {M.}~\bibnamefont
  {Ghazisaeidi}}, \bibinfo {author} {\bibfnamefont {L.~G.}\ \bibnamefont
  {Hector~Jr}}, \ and\ \bibinfo {author} {\bibfnamefont {W.}~\bibnamefont
  {Curtin}},\ }\href@noop {} {\bibfield  {journal} {\bibinfo  {journal}
  {Scripta Materialia}\ }\textbf {\bibinfo {volume} {75}},\ \bibinfo {pages}
  {42} (\bibinfo {year} {2014})}\BibitemShut {NoStop}%
\bibitem [{\citenamefont {Ready}\ \emph {et~al.}(2017)\citenamefont {Ready},
  \citenamefont {Haynes}, \citenamefont {Rugg},\ and\ \citenamefont
  {Sutton}}]{ready2017stacking}%
  \BibitemOpen
  \bibfield  {author} {\bibinfo {author} {\bibfnamefont {A.}~\bibnamefont
  {Ready}}, \bibinfo {author} {\bibfnamefont {P.}~\bibnamefont {Haynes}},
  \bibinfo {author} {\bibfnamefont {D.}~\bibnamefont {Rugg}}, \ and\ \bibinfo
  {author} {\bibfnamefont {A.}~\bibnamefont {Sutton}},\ }\href@noop {}
  {\bibfield  {journal} {\bibinfo  {journal} {Philosophical Magazine}\ }\textbf
  {\bibinfo {volume} {97}},\ \bibinfo {pages} {1129} (\bibinfo {year}
  {2017})}\BibitemShut {NoStop}%
\bibitem [{\citenamefont {Chaari}\ \emph {et~al.}(2014)\citenamefont {Chaari},
  \citenamefont {Clouet},\ and\ \citenamefont {Rodney}}]{chaari2014first}%
  \BibitemOpen
  \bibfield  {author} {\bibinfo {author} {\bibfnamefont {N.}~\bibnamefont
  {Chaari}}, \bibinfo {author} {\bibfnamefont {E.}~\bibnamefont {Clouet}}, \
  and\ \bibinfo {author} {\bibfnamefont {D.}~\bibnamefont {Rodney}},\
  }\href@noop {} {\bibfield  {journal} {\bibinfo  {journal} {Metallurgical and
  Materials Transactions A}\ }\textbf {\bibinfo {volume} {45}},\ \bibinfo
  {pages} {5898} (\bibinfo {year} {2014})}\BibitemShut {NoStop}%
\bibitem [{\citenamefont {Freitas}\ \emph {et~al.}(2016)\citenamefont
  {Freitas}, \citenamefont {Asta},\ and\ \citenamefont
  {De~Koning}}]{freitas2016nonequilibrium}%
  \BibitemOpen
  \bibfield  {author} {\bibinfo {author} {\bibfnamefont {R.}~\bibnamefont
  {Freitas}}, \bibinfo {author} {\bibfnamefont {M.}~\bibnamefont {Asta}}, \
  and\ \bibinfo {author} {\bibfnamefont {M.}~\bibnamefont {De~Koning}},\
  }\href@noop {} {\bibfield  {journal} {\bibinfo  {journal} {Computational
  Materials Science}\ }\textbf {\bibinfo {volume} {112}},\ \bibinfo {pages}
  {333} (\bibinfo {year} {2016})}\BibitemShut {NoStop}%
\bibitem [{\citenamefont {Lysogorskiy}\ \emph {et~al.}(2023)\citenamefont
  {Lysogorskiy}, \citenamefont {Bochkarev}, \citenamefont {Mrovec},\ and\
  \citenamefont {Drautz}}]{lysogorskiy2022active}%
  \BibitemOpen
  \bibfield  {author} {\bibinfo {author} {\bibfnamefont {Y.}~\bibnamefont
  {Lysogorskiy}}, \bibinfo {author} {\bibfnamefont {A.}~\bibnamefont
  {Bochkarev}}, \bibinfo {author} {\bibfnamefont {M.}~\bibnamefont {Mrovec}}, \
  and\ \bibinfo {author} {\bibfnamefont {R.}~\bibnamefont {Drautz}},\
  }\href@noop {} {\bibfield  {journal} {\bibinfo  {journal} {Physical Review
  Materials}\ }\textbf {\bibinfo {volume} {7}},\ \bibinfo {pages} {043801}
  (\bibinfo {year} {2023})}\BibitemShut {NoStop}%
\bibitem [{\citenamefont {Wentzcovitch}\ and\ \citenamefont
  {Cohen}(1988)}]{Wentzcovitch_PhysRevB.37.5571}%
  \BibitemOpen
  \bibfield  {author} {\bibinfo {author} {\bibfnamefont {R.~M.}\ \bibnamefont
  {Wentzcovitch}}\ and\ \bibinfo {author} {\bibfnamefont {M.~L.}\ \bibnamefont
  {Cohen}},\ }\href {\doibase 10.1103/PhysRevB.37.5571} {\bibfield  {journal}
  {\bibinfo  {journal} {Phys. Rev. B}\ }\textbf {\bibinfo {volume} {37}},\
  \bibinfo {pages} {5571} (\bibinfo {year} {1988})}\BibitemShut {NoStop}%
\end{thebibliography}
%

\end{document}